\documentclass[article,onecolumn,nofootinbib]{revtex4}
\usepackage{times}
\usepackage{natbib}
\usepackage{epsfig}
\usepackage{subfigure}
\usepackage{amsmath}
\usepackage{verbatim}
\usepackage[a4paper,left=1.2cm,right=1.2cm,bottom=1.5cm,top=3cm]{geometry}
\usepackage{hyperref}
\usepackage{color}

\newcommand{\beeq}{\begin{equation}}
\newcommand{\eneq}{\end{equation}}
\newcommand{\bear}{\begin{eqnarray}}
\newcommand{\enar}{\end{eqnarray}}

\newcommand{\bef}{\begin{figure}[!ht]}
\newcommand{\enf}{\end{figure}}

\newcommand{\fnl}{f_{NL}}
\newcommand{\kbf}{{\bf k}}
\newcommand{\xbf}{{\bf x}}

\newcommand{\tidal}{(k_x^2-k_y^2,2k_xk_y)}
\newcommand{\iacons}{\frac{C_1\rho_{\rm crit}\Omega_M}{D(z)} }

\newcommand{\SNfnlEUCLIDfour}{1.8}
\newcommand{\SNfnlEUCLIDjsfour}{2.5}
\newcommand{\SNbaoEUCLIDfour}{18}

\newcommand{\SNfnlEUCLID}{1.7}
\newcommand{\SNfnlEUCLIDjs}{2.1}
\newcommand{\SNbaoEUCLID}{15}

\newcommand{\SNfnldesi}{0.95}
\newcommand{\SNfnldesijs}{1.3}
\newcommand{\SNbaodesi}{12}

\newcommand{\SNbaocmass}{2.2}

\newcommand{\SNbaocmassTWELVE}{2.7}

\newcommand{\SNfnlboss}{0.14}
\newcommand{\SNfnlbossjs}{0.20}
\newcommand{\SNfnlbossTWELVE}{0.17}
\newcommand{\SNfnlbossjsTWELVE}{0.24}

\newcommand{\SNfnllowz}{$<$0.1}
\newcommand{\SNfnllowzjs}{0.26}
\newcommand{\SNfnllowzTWELVE}{0.11}
\newcommand{\SNfnllowzjsTWELVE}{0.33}
\newcommand{\SNbaolowz}{1.8}

\newcommand{\SNbaolowzTWELVE}{2.3}

\begin{document}

\title{Cosmological Information in the Intrinsic Alignments of Luminous Red Galaxies}
\author{Nora Elisa Chisari$^1$}
\altaffiliation{nchisari@astro.princeton.edu} 
\author{Cora Dvorkin$^2$}
\altaffiliation{cdvorkin@ias.edu}
\affiliation{$^1$ Department of Astrophysical Sciences, Princeton University, 4 Ivy Lane, Princeton, NJ 08544, USA}
\affiliation{$^2$ Institute for Advanced Study, School of Natural Sciences, Einstein Drive, Princeton, NJ 08540, USA}

\begin{abstract}
The intrinsic alignments of galaxies are usually regarded as a contaminant to weak gravitational lensing observables. 
The alignment of Luminous Red Galaxies, detected unambiguously in observations from the Sloan Digital Sky Survey, can be reproduced by the linear tidal alignment model of Catelan, Kamionkowski \& Blandford (2001) on large scales. In this work, we explore the cosmological information encoded in the intrinsic alignments of red galaxies. We make forecasts for the ability of current and future spectroscopic surveys to constrain local primordial non-Gaussianity and Baryon Acoustic Oscillations (BAO) in the cross-correlation function of intrinsic alignments and the galaxy density field. For the Baryon Oscillation Spectroscopic Survey, we find that the BAO signal in the intrinsic alignments is marginally significant with a signal-to-noise ratio of $\SNbaolowz$ and $\SNbaocmass$ with the current LOWZ and CMASS samples of galaxies, respectively, and increasing to $\SNbaolowzTWELVE$ and $\SNbaocmassTWELVE$ once the survey is completed. For the Dark Energy Spectroscopic Instrument and for a spectroscopic survey following the EUCLID redshift selection function, we find signal-to-noise ratios of $\SNbaodesi$ and $\SNbaoEUCLID$, respectively. Local type primordial non-Gaussianity, parametrized by $\fnl = 10$, is only marginally significant in the intrinsic alignments signal with signal-to-noise ratios $<2$ for the three surveys considered.
\end{abstract}

\maketitle

\section{Introduction}

The intrinsic alignments of galaxies are correlations between their positions, shapes and orientations that arise due to physical processes during their formation and evolution as tracers of the large-scale structure of the Universe. Examples of such processes are stretching by the tidal field of the large-scale structure \cite[][]{Catelan01,Hirata04,Hirata10} or clusters of galaxies \cite{Thompson76,Ciotti94}, the interaction between the angular momentum of a galaxy and the tidal torque \cite{Peebles69}, or dynamics along preferred directions \cite[][i.e., the filamentary distribution of galaxies]{Zhang09}.

The observed alignments of Luminous Red Galaxies (LRGs) \cite{Eisenstein01} have been shown to be adequately reproduced by the linear alignment (LA) model of \cite[][]{Catelan01,Hirata04,Hirata10} on large scales ($>10$Mpc$/h$) in the redshift range $0.16<z<0.47$ \cite{Okumura09,Blazek11}. On smaller scales, 
\cite{Hirata07,Bridle07} have suggested replacing the linear matter power spectrum by its non-linear analog. The non-linear alignment (NLA) model is widely used, despite the simplicity of the non-linear treatment, which does not evolve the non-linear scales consistently with the Poisson equation. A slight modification to the NLA model was proposed by \cite{Kirk12}, although once again without fully solving for the non-linear dynamics. \cite{Joachimi:2010xb} showed that the NLA model reproduces observations of the alignments of LRGs up to $z\sim 0.7$ and on scales of comoving projected separations $>6$ Mpc$/h$ resorting to the MegaZ-LRG sample \cite[][]{Collister07,Abdalla11}. On smaller scales, significant progress has been made in developing halo models that reproduce the currently available observational constraints and extend the forecasts of the intrinsic alignment contamination to weak lensing observables to higher redshifts and to satellite galaxies \cite{Schneider10,Joachimi13b,Joachimi13a}.
(Currently, there is no evidence of alignments for blue galaxies \cite{Mandelbaum11} and we do not consider them in this work.)

The intrinsic shapes and alignments of galaxies have been explored as a contaminant in weak gravitational lensing observables \citep{Hirata04,Joachimi10} but little is known about the cosmological information they encode. In the upcoming decades, imaging surveys like the Dark Energy Survey\footnote{http://www.darkenergysurvey.org} (DES) \cite{DES05}, Hyper-Suprime Cam\footnote{http://www.naoj.org/Projects/HSC/HSCProject.html} (HSC)\cite{Miyazaki12}, Pan-STARRS\footnote{http://pan-starrs.ifa.hawaii.edu/public/}, the Kilo-Degree Survey \footnote{http://kids.strw.leidenuniv.nl/} (KiDS), EUCLID\footnote{http://www.EUCLID-ec.org} \cite[]{EUCLIDRed}, the Large Scale Synoptic Telescope\footnote{http://www.lsst.org} (LSST) \cite{Ivezic08} and WFIRST\footnote{http://wfirst.gsfc.nasa.gov/} \cite{Spergel13} will map the large-scale structure of the Universe and explore the nature of dark energy \cite[][]{Weinberg12} by measuring the clustering of galaxies and the effect of weak gravitational lensing \cite[][]{Bartelmann01} over unprecedented volumes. 

In a complementary effort, ongoing spectroscopic surveys such as the Baryon Oscillation Spectroscopic Survey \cite{Dawson13}, and upcoming ones, such as SDSS-$IV$\footnote{http://www.sdss3.org/future/}, the Dark Energy Spectroscopic Instrument (DESI) \cite[]{DESIwhite}, EUCLID, WFIRST, and the Prime Focus Spectrograph (PFS) \cite[]{Takada12}, will gather spectra of millions of galaxies and quasars with the goal of measuring the distance scale probed by Baryon Acoustic Oscillations (BAO) and the growth of structure through redshift space distortions (RSD). 

In the context of these rich datasets, no probe of large-scale structure should be left unexplored. Intrinsic alignments are one such probe: an effect that was once just thought to be a contaminant to weak gravitational lensing measurements, could become a complementary source of cosmological information. In the LA model, the tidal field determines the strength and evolution of the alignments, which depend on the growth function of the matter density perturbations and their power spectrum. In particular, the LA prediction is thus sensitive to RSD, primordial non-Gaussianity and BAO. Another example of the potential of alignments to constrain cosmology was proposed by \cite{Schmidt12}, who suggested that gravitational waves from inflation could be detected in the intrinsic alignments of galaxies. 

In this work we assume that red galaxies (the ancestors of low redshift LRGs, and which we will henceforth also refer to as LRGs) follow the NLA model and we study the cosmological information contained in the galaxy density-intrinsic shear cross-correlation function. In Section \ref{sec:fidu}, we summarize the tidal alignment model, we construct the galaxy density-intrinsic shear cross-correlation and we discuss possible contaminants and uncertainties of the model. In Section \ref{sec:smooth} we study the typical scales at which the tidal field has an impact on the alignment of a galaxy. In Section \ref{sec:redshift} we explore an alternative to the NLA model in which the alignment between a galaxy and the tidal field of the large-scale structure occurs instantaneously. In Section \ref{sec:cosmo}, we study different cosmological observables that can be constrained using the intrinsic ellipticity-galaxy density cross-correlation: primordial non-Gaussianity (Section \ref{sec:fnl}) and the BAO (Section \ref{sec:bao}). In Section \ref{sec:fnl}, we also analyse the effect of RSD on the cross-correlation. In Section \ref{sec:galaxy-galaxy} we compare our results to the constraints the come from galaxy-galaxy lensing. In the Appendix we give a derivation of the Gaussian covariance matrix of the galaxy density-intrinsic shear cross-correlation in real space.

Unless otherwise noted, throughout this paper we work with the following {\it Planck} \cite{Ade:2013zuv} fiducial cosmology: $\Omega_{\rm b}h^2=0.022$, $\Omega_{\rm CDM}h^2=0.1204$, $h=0.67$, $\Omega_K=0$, $A_s=2.21\times10^{-9}$, $n_s=0.9619$, $k_p=0.05$ Mpc$^{-1}$ and we define $\Omega_M=\Omega_b+\Omega_{\rm CDM}$. 

\section{The tidal alignment model}
\label{sec:fidu}

We define the ellipticity of a galaxy as $e=(1-q^2)/(1+q^2)$, where $q$ is the ratio of the minor axis to the major axis of the best-fit ellipse to the galaxy image. The ellipticity can be decomposed in two components, $e_+ =e \cos(2\theta)$ and $e_\times= e \sin(2\theta)$, with $\theta$ the position angle of the galaxy. The component $e_+$ indicates radial (if negative) or tangential (if positive) alignment of a galaxy with respect to another galaxy. The $e_\times$ component measures the $45\deg$ rotation with respect to $e_+$ \cite{Bernstein02}. The distortion, $\gamma$, acting on a galaxy is the change in its ellipticity; it can also be decomposed into $\gamma_+$ and $\gamma_\times$. The relation between distortion and ellipticity is $\gamma=e/2\mathcal{R}$, where $\mathcal{R}$ is the responsivity factor, the response of the ellipticity of a galaxy to an applied distortion \cite{Bernstein02}. 

When we measure the shapes of galaxies, we are measuring a combination of the effect of the tidal field $\gamma^I$, the effect
of weak gravitational lensing on those shapes $\gamma^G$, and the intrinsic random shapes, $\gamma^{\rm rnd}$:

\beeq
\gamma^{\rm obs}=\gamma^I+\gamma^G+\gamma^{\rm rnd}.
\label{eq:allellip}
\eneq

The topic of this work is the correlation of $\gamma^I$ with the galaxy field. We will consider as the fiducial model of intrinsic alignments the NLA model proposed by \cite{Bridle07}. In this model, an intrinsic shear due to the tidal field of the primordial potential, $\phi_p$, 

\beeq
\gamma^I(\xbf,z) \equiv (\gamma_+^I,\gamma_\times^I) = -\frac{C_1}{4\pi G}(\nabla_x^2-\nabla_y^2,2\nabla_x\nabla_y)\phi_p(\xbf)
\label{eqn_hirata_gammaI}
\eneq

\noindent acts on a galaxy, where $C_1$ is some undetermined constant measuring the strength of the alignment, $G$ is Newton's gravitational constant and the quantity in parentheses is the tidal tensor operator on the plane of the sky. 

The primordial potential is evaluated at a redshift $z_p$, during matter domination, when the galaxy was formed. This is an $ansatz$, since there is no first principle model from which the LA model is derived. The redshift at which intrinsic alignments are set is unconstrained, but \cite{Joachimi:2010xb} have shown that the current measurements of the intrinsic alignments of LRGs are consistent with the primordial alignment model.

On large scales, the density field behaves linearly, $\delta_{\rm lin}$, and can be related to the primordial potential through the Poisson equation,

\beeq
\phi_p(\kbf) = - \frac{4\pi Ga^3\bar{\rho}(z)}{D(z)}\frac{\delta_{\rm lin}(\kbf,z)}{k^2}, 
\label{eqn:poisson}
\eneq

\noindent in Fourier space, where $\bar{\rho}(z)$ is the mean density of the Universe at redshift $z$, $D(z)$ is the growth function and $a$ is the scale factor.  

In practice, we can only measure shapes of galaxies at the positions of galaxies, hence, we observe the density-weighted intrinsic shear
field, $\tilde{\gamma}^I = (1+\delta_g)\gamma^I$, with $\delta_g=b\delta_{\rm lin}$ the galaxy overdensity field and $b$, a scale-independent bias. \footnote{For simplicity, we consider the galaxy bias to be independent of luminosity. The change of the bias with luminosity will depend on the properties of the galaxies selected. We do not attempt to model this effect in this work, but the bias as a function of redshift and luminosity can be constrained by the galaxy auto-correlation function. Moreover, there is significant evidence that the strength of the alignments increases with luminosity \cite{Hirata07,Joachimi:2010xb}, although this could also be consequence of mass dependence, as suggested by \cite{Joachimi13a}.} This weighting is of particular importance for intrinsic alignments, since galaxies that enter the correlation are physically associated. The details of the weighting might also depend on selection effects of the galaxy sample, such as flux $S/N$ or apparent size, which we do not model in this work. In the linear regime, this implies an effective rescaling of the bias, while in the non-linear regime, it can change the shape of the smoothing filter. A similar source-lens clustering is negligible in the context of galaxy-galaxy lensing, even at the typical current precision of photometric redshifts \cite{Schmidt09}. 

The intrinsic shear in Fourier space is given by

\beeq
\gamma^I(\kbf,z) = \frac{C_1}{4\pi G}(k_x^2-k_y^2,2k_xk_y)\mathcal{S}[\phi_p(\kbf)],
\label{eqn:hirata_gammaI}
\eneq

\noindent where we have defined a smoothing filter for the primordial potential, $\mathcal{S}$, that removes the effect of the tidal field on scales smaller
than the typical halo inhabited by LRGs. The purpose of applying a smoothing filter is to smooth the tidal field within the scale of the halo inhabited by the galaxy and to suppress the correlation due to non-linear effects on small scales that perturb the alignment. Numerically, the smoothing filter also avoids spurious features in the correlation function due to a sharp cut-off in $k-$space. We discuss the effect of the smoothing filter in Section \ref{sec:smooth}.

The density-weighted intrinsic shear field is given by a convolution in Fourier space,

\beeq
\tilde{\gamma}^I(\kbf,z) = \int d^3\kbf_1 \gamma^I(\kbf-\kbf_1,z)\left[\delta^{(3)}(\kbf_1)+\frac{b}{(2\pi)^3}\,\delta_{\rm lin}(\kbf_1,z)\right],
\label{eq:observed_shape}
\eneq

\noindent where $\delta^{(3)}$ is the three dimensional Dirac delta.

The galaxy density field and the $+$ component of the weighted intrinsic alignment tensor are correlated with a power spectrum given by

\beeq
P_{g+}(\kbf,z) = b \iacons \frac{k_x^2-k_y^2}{k^2} P_{\delta}^{\rm lin}(\kbf,z).
\label{eq:fiduc_gI}
\eneq

In the NLA model, $P_{\delta}^{\rm lin}(\kbf,z)$ is replaced by the non-linear matter power spectrum in Eq. \eqref{eq:fiduc_gI}.  \cite{Kirk12} suggest a slightly modified version of the NLA model, where $g+$ is constructed assuming that the tidal field does not undergo non-linear evolution, while $\delta_g$ does. This is also an approximation to the problem of correlating the non-linear density field with the primordial tidal field, but does not improve on the physical treatment of the non-linear dynamics. In this work we use the code CAMB\cite{Lewis:2002ah} to obtain the non-linear matter power spectrum.

\subsection{Modeling of the correlations}
\label{sec:correlation}

For modeling the correlations between galaxies and intrinsic shapes, which is the subject of this work, we will consider a redshift-space correlation function between two observables $a$ and $b$,

\beeq
\xi_{ab}({\bf r_p},\Pi,z)= \langle a({\bf 0},\chi(z),z) b({\bf r_p},\chi(z)+\Pi,z) \rangle ,
\label{eq:basicxi}
\eneq

\noindent where ${\bf r_p}$ is the comoving projected separation vector, in the plane of the sky, $\chi$ is the comoving distance along the line of sight,
and $\Pi$ is the perpendicular component of the separation vector projected along the line of sight, and $z$ is the redshift. 

We can relate the correlation function in redshift space to the power spectrum of the
two observables $a$ and $b$ through:

\beeq
\xi_{ab}({\bf r_p},\Pi,z)= \int \frac{d^2{\bf k_{\perp}} dk_z}{(2\pi)^3} P_{ab}(\kbf,z) e^{i({\bf k_{\perp}} \cdot {\bf r_p}+k_z\Pi)}.
\label{eq:defcorrel}
\eneq

\noindent
Analogously to the choice of cylindrical coordinates in Eq. (\ref{eq:basicxi}), we choose cylindrical coordinates in Fourier space (along the line of sight, $k_z$ and perpendicular to it, $k_{\perp}$). We will henceforth refer to $\xi_{ab}(r_p,\Pi,z)$ as the angular average over the directions of ${\bf r_p}$ (i.e., on the plane of the sky) of Eq. \eqref{eq:basicxi}. The projected correlation function under the Limber approximation is defined as

\beeq
w_{ab}(r_p)=\int dz \mathcal{W}(z) \int_{-\Pi_{\rm max}}^{\Pi_{\rm max}} d\Pi \,\xi_{ab}(r_p,\Pi,z),
\label{eq:defcorrelint}
\eneq

\noindent where the weight function $\mathcal{W}(z)$ depends on the observables. The advantage of the estimator in Eq. \eqref{eq:defcorrelint} is that it only computes the correlation between galaxies in a box of length $2\Pi_{\rm max}$ along the line of sight. This procedure reduces the contamination from other sources of correlation, as we will discuss in Section \ref{sec:contam}. We apply here the weighting derived in \cite{Mandelbaum11}:

\beeq
\mathcal{W}(z) =\frac{p^2(z)}{\chi^2(z)\frac{d\chi}{dz}(z)}\left[ \int dz \frac{p^2(z)}{\chi^2(z)\frac{d\chi}{dz}(z)} \right]^{-1},
\eneq

\noindent where $p(z)$ is the redshift distribution of the galaxies in the sample, normalized to unity.
This is the correct weighting when counting pairs of galaxies in the cylinder defined by coordinates $(r_p,\Pi)$. The comoving distance factors account for the change of comoving volume with redshift.

\subsection{Intrinsic shape correlations}
\label{sec:observed}

In the tidal alignment model, $\gamma^I$ depends on the primordial gravitational potential through Eq. (\ref{eqn_hirata_gammaI}). As a consequence, $\gamma^I$ and the density field are correlated. This correlation can be directly measured, by taking galaxies as tracers of the density field and building the correlation function with their measured ellipticities. We will refer to the correlations between LRG positions and LRG shapes as $gI$. We will study the cosmological information imprinted on the correlations between the intrinsic ellipticities and the density field, as traced by galaxies.
Under these assumptions, the cross-correlation between the observed density field and the observed shapes, given by Eq. (\ref{eq:observed_shape}), at a given redshift is

\beeq
gI = b(z)\langle\delta\gamma^I\rangle,
\label{eq:corrobs}
\eneq

\noindent We have neglected here the contributions of galaxy-galaxy lensing and magnification bias to the correlation. We will discuss those contributions in Section \ref{sec:contam}.


In the tidal alignment model, the $gI$ and $II$ correlation functions were computed by \cite{Hirata04,Hirata10}. We incorporate the effect of RSD on large scales by transforming the non-linear matter power spectrum, $P_{\delta}(k,z)$ to the anisotropic redshift-space distorted power spectrum, $P_s(\kbf,z)$, through the transformation derived by \cite{Kaiser87} and \cite{Hamilton92} and applied to intrinsic alignments already in \cite{Blazek11}. Moreover, because of the smoothing filter acting on the primordial potential in Eq. (\ref{eqn:hirata_gammaI}), the tidal field power spectrum is further multiplied by the smoothing filter, 

\beeq
P_s(\kbf,z)=P_{\delta}(k,z)\mathcal{S}(k)\left(1+\beta\mu^2\right)^2,
\label{eq:rsdsm}
\eneq

\noindent where $\beta=\Omega_m^{0.55}(z)/b(z)$ \cite{Linder:2005in}, and $\mu=\cos(\theta_k)=k_z/k$ where $\theta_k$ is the angle between $\kbf$ and the line of sight. This approximation to the effect of RSD has been shown to be valid over an arbitrary range of scales \cite{Yoo13}.

The redshift space $gI$ correlation functions, projected along the line of sight on the cylinder defined by $\Pi_{\rm max}$, are given by:

\bear
\xi_{g+} (r_p,z) &=& \frac{b}{\pi^2} \iacons \int_0^\infty dk_z\int_0^\infty dk_{\perp} \frac{k_{\perp}^3}{k^2k_z}P_s(\kbf,z)\sin(k_z\Pi_{\rm max})J_2(k_{\perp} r_p),
\enar

\noindent where $J_2$ is the second order spherical Bessel function, and $\xi_{g\times}(r_p,\Pi,z)=0$. Integrating $\xi_{g+}$ over the redshift range of the LRG sample using the Limber approximation, we obtain:

\beeq
w_{g+}(r_p) = \int dz \mathcal{W}(z)  \frac{b}{\pi^2} {C_1\rho_{\rm crit}\Omega_M\over D(z)} \int_0^\infty dk_z\int_0^\infty dk_{\perp} \frac{k_{\perp}^3}{(k_{\perp}^2+k_z^2)k_z}P_s(\kbf,z)\sin(k_z\Pi_{\rm max})J_2(k_{\perp} r_p).
\label{eq:wgp}
\eneq

\noindent We have chosen $\Pi_{\rm max}=80\,$Mpc$/h$ in agreement with \cite{Blazek11}.

For computing the auto-correlation functions between the different components of the ellipticity, $++$ and $\times\times$, we find it convenient to define the following functions,

\bear
f_E(\kbf)&=&\frac{k_x^2-k_y^2}{k^2},\nonumber\\
f_B(\kbf)&=&\frac{2k_xk_y}{k^2}.
\enar

\noindent With this notation, the auto-power spectrum of $\gamma^I_+$ is given by

\bear
P_{++}(\kbf,z) &=& \left( \iacons\right)^2 P_s(\kbf,z)\mathcal{S}(k)f_E^2(\kbf) \nonumber\\
&+& \left( b\iacons\right)^2 \int \frac{d^3\kbf_1}{(2\pi)^3} f_E(\kbf_1)P_s(\kbf_1,z) P_s(|\kbf-\kbf_1|,z)\mathcal{S}(k_1)\mathcal{S}(|\kbf-\kbf_1|)\nonumber\\
&\times&\left\{ f_E(\kbf_1)  +f_E(\kbf-\kbf_1) \right\} .
\label{eq:pppower}
\enar

\noindent For obtaining $P_{\times\times}(\kbf,z)$, it suffices to replace $f_E$ by $f_B$ in the above equation 

The smoothing filter has a larger impact on the auto-power spectra of the intrinsic ellipticities, $P_{++}$ and $P_{\times\times}$, since the smoothing filter appears squared in Eq. (\ref{eq:pppower}). From now on, we neglect quadratic terms in the power spectrum in Eq. (\ref{eq:pppower}), since they do not have a significant contribution to $P_{++}(k)$ \cite{Hirata04}. 

The auto-correlation functions of intrinsic ellipticities are given by 

\bear
w_{++}(r_p)&=& \frac{1}{2\pi^2}\int dz \mathcal{W}(z) \left(\frac{C_1\rho_{\rm crit}\Omega_M}{D(z)}\right)^2\int_0^\infty dk_z\int_0^\infty dk_{\perp} \frac{k_{\perp}^5}{(k_{\perp}^2+k_z^2)^2k_z} P_s(\kbf,z)\mathcal{S}(k)\nonumber\\
&\times&\sin(k_z\Pi_{\rm max})[J_0(k_{\perp} r_p)+J_4(k_{\perp} r_p)],
\label{eqn_correl_plusplus}
\enar

\bear
w_{\times\times}(r_p) &=& \frac{1}{2\pi^2}\int dz \mathcal{W}(z) \left(\frac{C_1\rho_{\rm crit}\Omega_M}{D(z)}\right)^2\int_0^\infty dk_z\int_0^\infty dk_{\perp} \frac{k_{\perp}^5}{(k_{\perp}^2+k_z^2)^2k_z}P_s(\kbf,z)\mathcal{S}(k)\nonumber\\
&\times&\sin(k_z\Pi_{\rm max})[J_0(k_{\perp} r_p)-J_4(k_{\perp} r_p)].
\label{eqn_correl_crosscross}
\enar

\subsection{Strength of alignment}
\label{sec:fitC1}

We obtain $C_1$ from fitting the density-intrinsic shear cross-correlation. As illustrated by Eq. (\ref{eq:allellip}), the shape of a galaxy has contributions from the shear, the tidal field and a random noise component. For low redshift galaxies, the effect of gravitational lensing is negligible. We use low redshift observations of the intrinsic shapes of LRGs by \cite[][]{Okumura09} to constrain the value of $C_1$, the amplitude of the intrinsic alignment effect. We assume $b=2.12$, the bias derived for this sample of LRGs from their clustering \cite{Blazek11}. 

\cite[][]{Okumura09} measured $w_{g+}(r_p)$ for $73,935$ LRGs in SDSS DR6 in the redshift range of $0.16<z<0.47$ and with a median redshift of $\bar{z}\sim0.32$. In that work, LRGs are selected by applying the technique of \cite{Reid09}, identifying central LRGs and removing satellites for each halo. The volume associated with each halo is a cylinder with dimensions $r_p=0.8$Mpc$/h$ and $\Delta\Pi=20$Mpc$/h$. This cut effectively removes the one-halo term contribution to $w_{g+}$.

We perform a least-squares fit to $w_{g+}(r_p)$ and obtain $C_1\rho_{\rm crit}=0.131\pm0.013$ with a reduced $\chi^2=3.2$, for $r_p>1$ Mpc$/$h. Figure \ref{fig:fitC1} shows the results of our fit to the data from \cite[][]{Okumura09} (their figure 3). If we only take into account points at $r_p>10$ Mpc$/h$ for the fit, we obtain a consistent result of $C_1\rho_{\rm crit}=0.126\pm0.013$ with a reduced $\chi^2=0.9$. \cite[]{Blazek11} also performed the fit on large scales, where the effect of the smoothing filter is not significant; our results are in agreement with theirs. Because we are interested in cosmological constraints on large scales, we adopt the latter value of $C_1$ as our fiducial amplitude for the remaining sections. Our numerical convergence tests for $w_{g+}$ indicate a $1.2\%$ additional uncertainty in the value of $C_1$. 

We can compare the scatter in intrinsic ellipticities to the predicted value from the NLA model, which is obtained from the power spectrum of the gravitational potential as in Eq. (8) of  \cite{Catelan01}: 

\bear
\langle e^2 \rangle &=& 4\mathcal{R}^2 \langle \gamma^{I\,2}_+ + \gamma^{I\,2}_\times \rangle ,\nonumber\\
&=& 4\mathcal{R}^2\int_{z_{\rm min}}^{z_{\rm max}} dz \mathcal{W}(z) \left( \iacons\right)^2\int \frac{d^3\kbf}{(2\pi)^3} P_s(\kbf,z)  \left[ f_E^2(\kbf)+f_B^2(\kbf) \right]  \mathcal{S}(k),\nonumber\\
& \approx& 2 \times 10^{-3}.
\label{eqn:ermsorig}
\enar

Integrated over the range $0.16<z<0.47$, the NLA prediction yields a much lower value for $\langle e^2 \rangle$ than the measured value for the LRGs of \cite{Okumura09}, for which $\langle e_{\rm obs}^2 \rangle \approx 0.1$ ($\mathcal{R}=0.947$). We are assuming here that $\mathcal{R}$ is independent of redshift. In practice, this will most likely not be true, but it would require an unrealistic variation of at least an order of magnitude in $\mathcal{R}$ to bring the NLA model in agreement with $\langle e_{\rm obs}^2 \rangle$. The shapes used by \cite{Okumura09} are not corrected for the Point Spread Function (the response of an unresolved source to the combined effect of the telescope optics and the atmosphere). This results in an overestimation of $\langle e^2 \rangle$ but, once more, this effect, although highly significant \cite{Blazek11}, is not sufficient to account for the difference between the NLA prediction and the observed value. ``Shape noise'' due to a random component of ellipticity, $\gamma_{\rm rnd}$, is the dominant contribution to the scatter and prevents us from estimating $C_1$ with this method.

\bef
\centering
\includegraphics[width=0.7\textwidth]{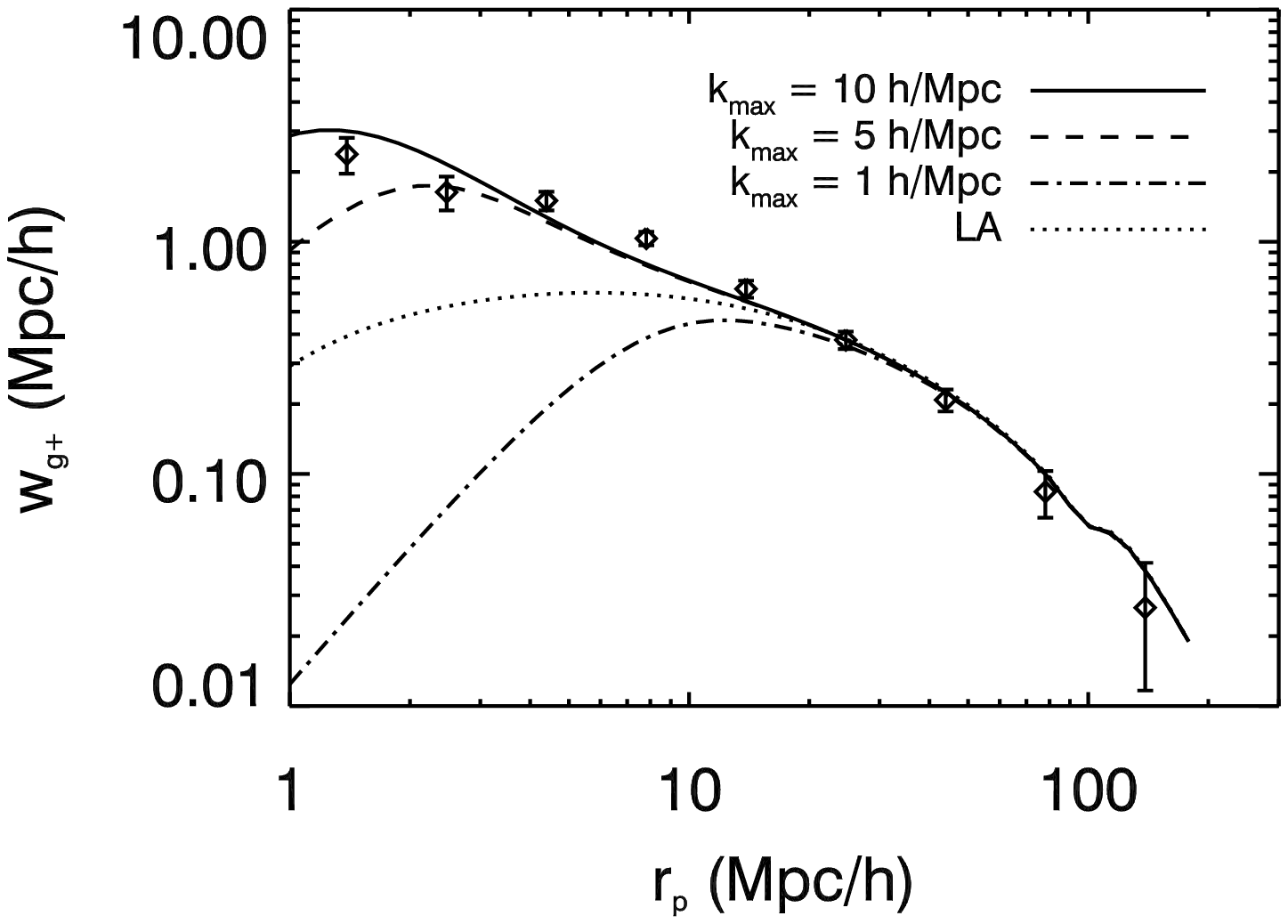}
\caption{The observed galaxy-intrinsic ellipticity correlation function (diamonds), $w_{g+}(r_p)$, from  \cite{Okumura09} compared
to our best fit intrinsic alignment model with a smoothing filter of typical scale $k_{\rm max} = 10\,h$/Mpc (solid line).
The dashed and dotted-dashed lines correspond to the predictions of the intrinsic alignment model for the cases with $k_{\rm max} = 5\,h/$Mpc and $k_{\rm max} = 1\,h/$Mpc, respectively (using 
the best fit $C_1$ from the $k_{\rm max} = 10\,h/$Mpc case).
The dotted line corresponds to the linear alignment model (which uses the linear matter power spectrum).}
\label{fig:fitC1}
\enf

\subsection{Smoothing filter}
\label{sec:smooth}

\cite{Hirata04} suggest the use of a top-hat filter in $k-$space in Eq. (\ref{eqn:hirata_gammaI}) to remove the effect of the tidal field in galactic scales. \cite{Catelan01} use a top-hat filter in real space. \cite{Blazek11} do not use any particular smoothing because they focus on scales $>10$ Mpc$/h$. These assumptions are all consistent with the assumption of linearity. In this work we use a well-behaved filter in $k$-space in order to avoid non-physical behavior of real-space correlation functions. We thus consider the following smoothing filter as our fiducial filter:

\beeq
\mathcal{S}(k) = \exp[-(3k/k_{\rm max})^2].
\label{eqn:smooth}
\eneq

\noindent The filter is a Gaussian function that decays by a factor $1/e$ at $k=k_{\rm max}/3$. This choice of smoothing filter removes spurious oscillations in the correlation function produced by a sharp cut-off in $k$-space.

We can see in Figure \ref{fig:fitC1} that the smoothing considerably changes the predicted correlation function. The smoothing filter that best reproduces the measured correlation function has $k_{\rm max} = 10h/$Mpc. This is roughly consistent with the typical scale of an LRG halo found by \cite{Reid09}, of $r_{\rm halo}= 0.8$ Mpc$/$h. The linear alignment model curve (with a filter of $k_{\rm max} = 10h/$Mpc) lies between the non-linear alignment models with $k_{\rm max} = 1 h/$ Mpc and $k_{\rm max} = 5 h/$Mpc. This is because, even though it does not capture the non-linear physics well enough, the linear alignment model still has non-zero power on small scales. Removing the smoothing filter for the LA model leads to changes of $<0.5\%$ on scales $>10$Mpc$/h$, which are negligible compared to the effect of non-Gaussianity or the BAO in $w_{g+}$ in Section \ref{sec:cosmo}.

\bef
\centering
\subfigure[\hskip 2pt The auto-correlation function of $\gamma_+$
predicted for the sample of \cite{Okumura09} for smoothing
filters of different $k_{\rm max}$ values.]{
\includegraphics[width=0.45\textwidth]{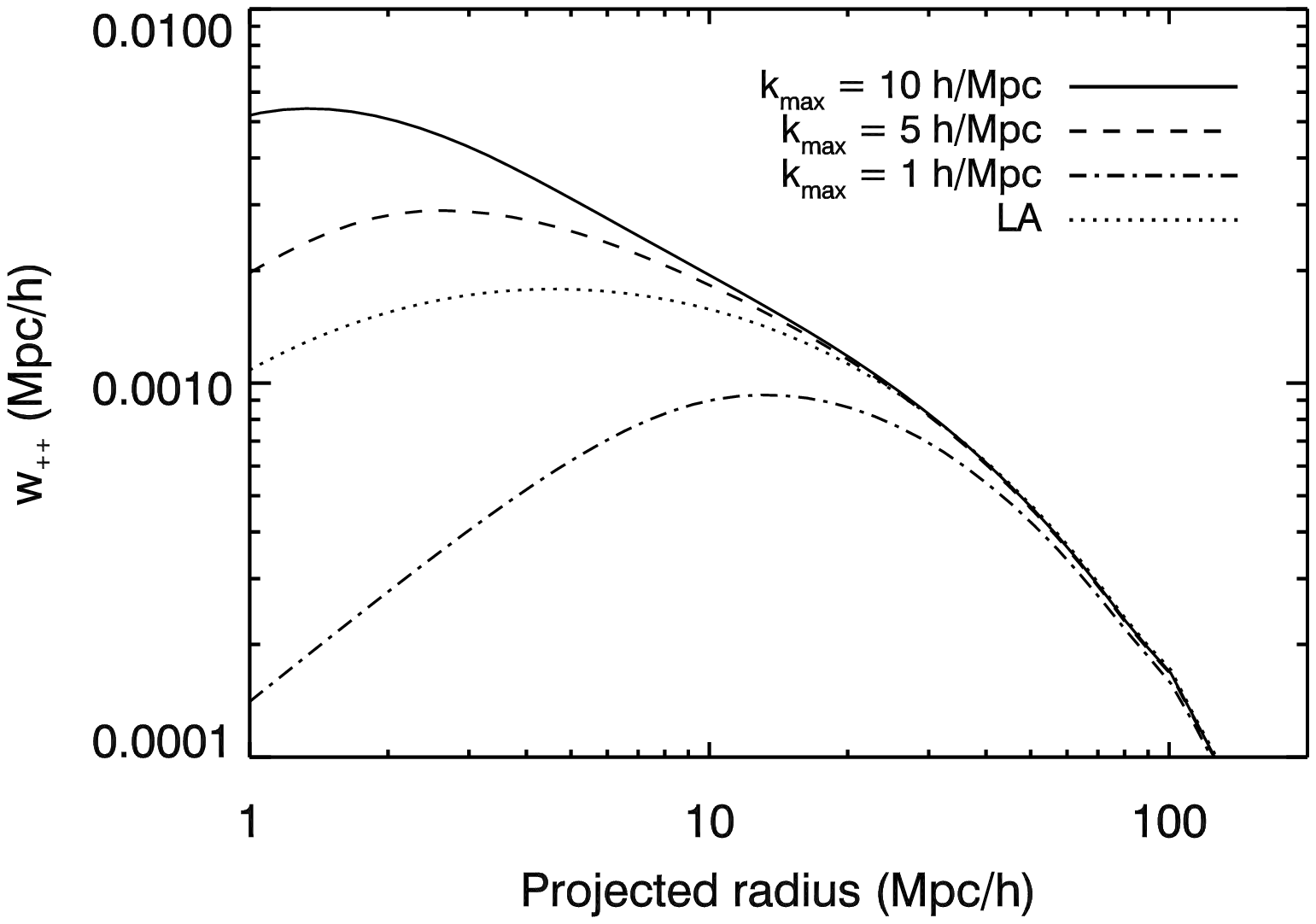}
}
\subfigure[\hskip 2pt Same as (a) for $\gamma_\times$]{
\includegraphics[width=0.45\textwidth]{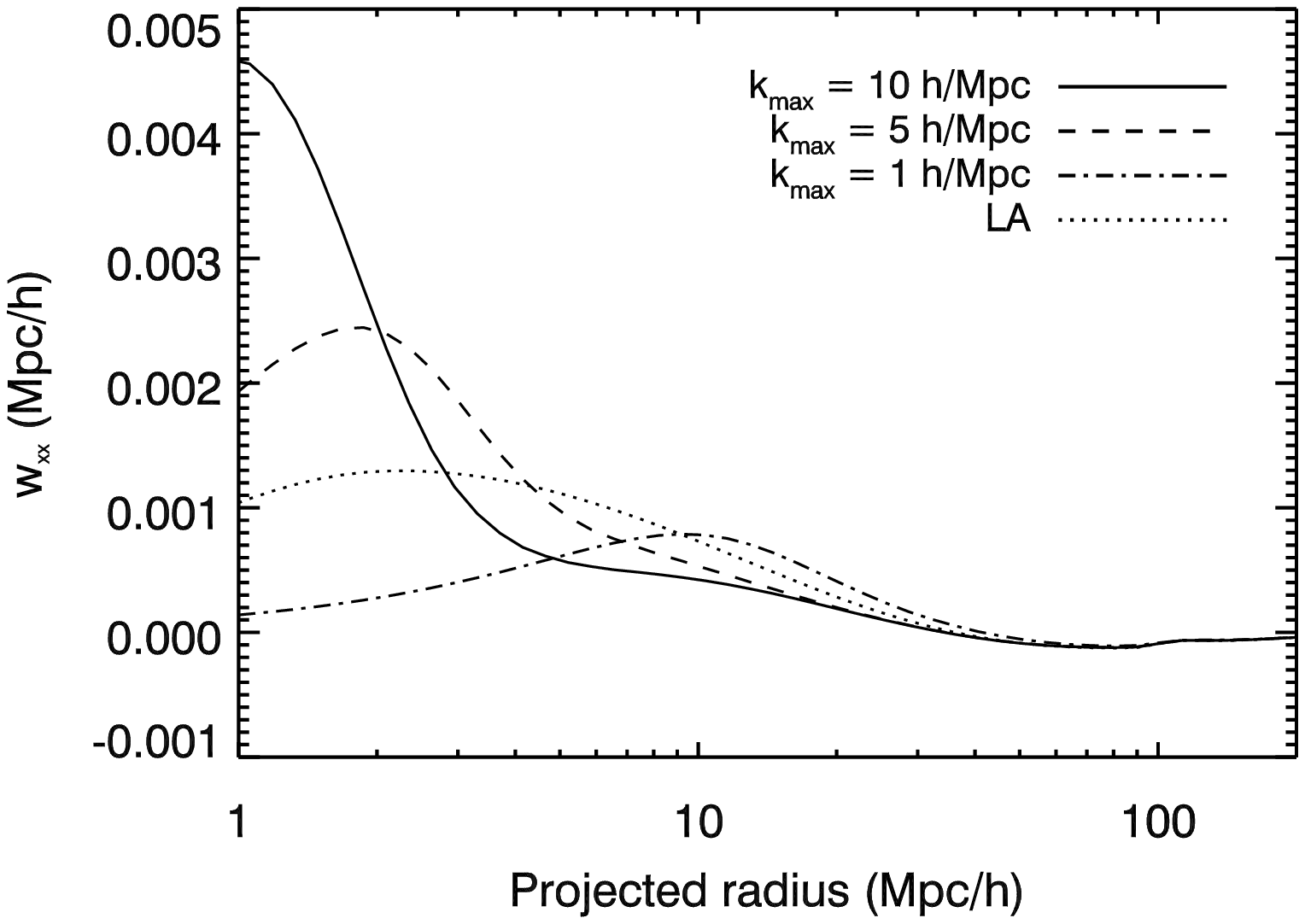}
}
\caption{The auto-correlation functions of the intrinsic ellipticity
components: $\gamma^I_+$ (left) and $\gamma^I_\times$ (right) predicted for the sample of \cite{Okumura09} for smoothing
filters of different $k_{\rm max}$ values. The LA curve is subject to smoothing with $k_{\rm max}=10h/$Mpc.}
\label{fig:autoII}
\enf

The smoothing filter also has a noticeable effect on the auto-correlation function of intrinsic shapes, $II$. In Figure \ref{fig:autoII}, we show the predicted $II$ correlation for the sample of \cite{Okumura09}. For $w_{++}(r_p)$ and $w_{\times\times}(r_p)$, the effect of the smoothing filter is smaller on small scales in comparison to $w_{g+}(r_p)$, but it is still present at scales of $20$ Mpc$/h$. As in Section \ref{sec:fitC1}, we have neglected the quadratic contributions of the matter fluctuation power spectrum in this section. The auto-correlation functions of LRG shapes at $z<0.5$ were measured by \cite[]{Okumura09b} and shown to be in agreement with the LA model on large scales by \cite[]{Blazek11}. For the sample of \cite{Okumura09b}, $w_{\times \times}(r_p)$ is consistent with $0$ at $r_p<10$Mpc$/h$, where the effect of the smoothing filter is large.

\subsection{When were intrinsic alignments imprinted?}
\label{sec:redshift}

In the study by \cite{Hirata04,Hirata10}, intrinsic alignments are imprinted at the redshift of formation of the galaxy, $z_p$. An alternative to the primordial alignment model is to assume that there is an instantaneous response of the galaxy shape to the tidal field. These two models result in different redshift dependence of the amplitude of the intrinsic alignment signal. In the case of instantaneous alignment, Eq. (\ref{eqn:hirata_gammaI}) is modified to be

\beeq
\gamma^I(\kbf,z) = \frac{C_1}{4\pi G}(k_x^2-k_y^2,2k_xk_y)\mathcal{S}[\phi(\kbf,z)],
\label{eqn:ia_instant}
\eneq

\noindent where $\phi$ is the gravitational potential at redshift $z$. Any model in which the galaxy shape and orientation is determined between the redshift of formation of the galaxy and the redshift of observation will have an amplitude that lies between the instantaneous and the primordial alignment case. The instantaneous and primordial alignment model differ from one another by a factor $D^{-1}(z)$, the inverse of the growth function. 

The constraints on the amplitude of the alignment signal obtained by \cite{Hirata07} and \cite{Joachimi:2010xb} are inconclusive regarding the redshift dependence of the alignment signal. \cite{Hirata07} measured the $GI$ correlation of LRGs in the SDSS DR4 sample ($0.16<z<0.35$) and in the 2SLAQ LRG sample ($0.4<z<0.8$) \cite[]{Cannon06}. Their results are consistent with the LA model, albeit with large error bars. Similarly, in \cite{Joachimi:2010xb}, the authors study the redshift dependence of $w_{g+}$ by combining previous results for the LRGs of the SDSS DR4 by \cite{Hirata07} with the MegaZ-LRG sample (at $\bar{z}\sim 0.5$) \cite[][]{Collister07,Abdalla11}, selected from SDSS DR6. They find no evidence for a redshift dependence other than the one proposed by the LA model. More specifically, in their Table $3$, they show their constraints on the amplitude of the $gI$ correlation for the MegaZ-LRG sample in two redshift bins with median redshifts of $z=0.49$ and $z=0.59$. The $gI$ amplitude is quoted relative to a fiducial value of $C_1$ and correcting for galaxy-galaxy lensing and magnification bias contaminations. At $1\sigma$, it is not possible to distinguish between the instantaneous and the primordial models for those redshift bins. A halo model approach to intrinsic alignments, applied to the Millenium Simulation, displays an evolution consistent with the LA model as well \cite{Joachimi13a}.

In Figure \ref{fig:redshiftdep}, we explore the ability of future surveys to distinguish between the instantaneous and the primordial LA model. Having normalized the amplitude of the intrinsic alignment signal to unity today, we show the ratio between the instantaneous and the primordial alignment models for the $gI$ correlation and the $II$ auto-correlation functions. This ratio is proportional to the growth function, $D(z)$, for $gI$, and to $D^2(z)$ for $II$. The instantaneous model has a steeper dependence with redshift. While this difference is really a function of scale, since the Kaiser factor of Eq. \eqref{eq:rsdsm} for RSD depends on redshift, this affects $w_{g+}$ very similarly for the surveys considered in this work, as will be seen in Section \ref{sec:fnl} and Figure \ref{fig:rsd}, such that we only consider an overall variation in amplitude of the IA signal with redshift in this section. 
(Notice that in the non-linear regime, the details of the smoothing filter and the non-linear approximation adopted \cite{Bridle07,Kirk12} have a significant impact on the evolution of alignment signal.)

The percentual difference in the amplitude of the $gI$ alignment signal in the primordial and the instantaneous model is $\sim20\%$ at $z=0.5$ and reaches $35\%$ at the median redshift of EUCLID. If we had assumed that the correlation length of galaxies is fixed, the bias would evolve proportionally to the growth function. An instantaneous LA model with constant bias has a similar redshift dependence to a primordial model with $b\propto D(z)$. This result stresses the necessity of measuring the bias of galaxies as a function of redshift, from the auto-correlation function of the positions of galaxies, complementarily to the intrinsic alignment correlation in order to distinguish between these models. While the $II$ dependence is steeper, this signal is also harder to measure due to shape measurement systematics.

Although the LA model reproduces the LRG alignments at $z<0.5$, it does not make a prediction for the strength of alignment, i.e., the value of $C_1$. The coupling of the baryons to the tidal field of the dark matter will ultimately determine the value of $C_1$ and the effective redshift of the tidal field. A possibility that we have not explored here is that there is a lag between the tidal field and the response of the galaxy. These variants to the LA model can only be fully studied with hydrodynamical numerical simulations of galaxy formation.

\bef
\centering
\includegraphics[width=0.5\textwidth]{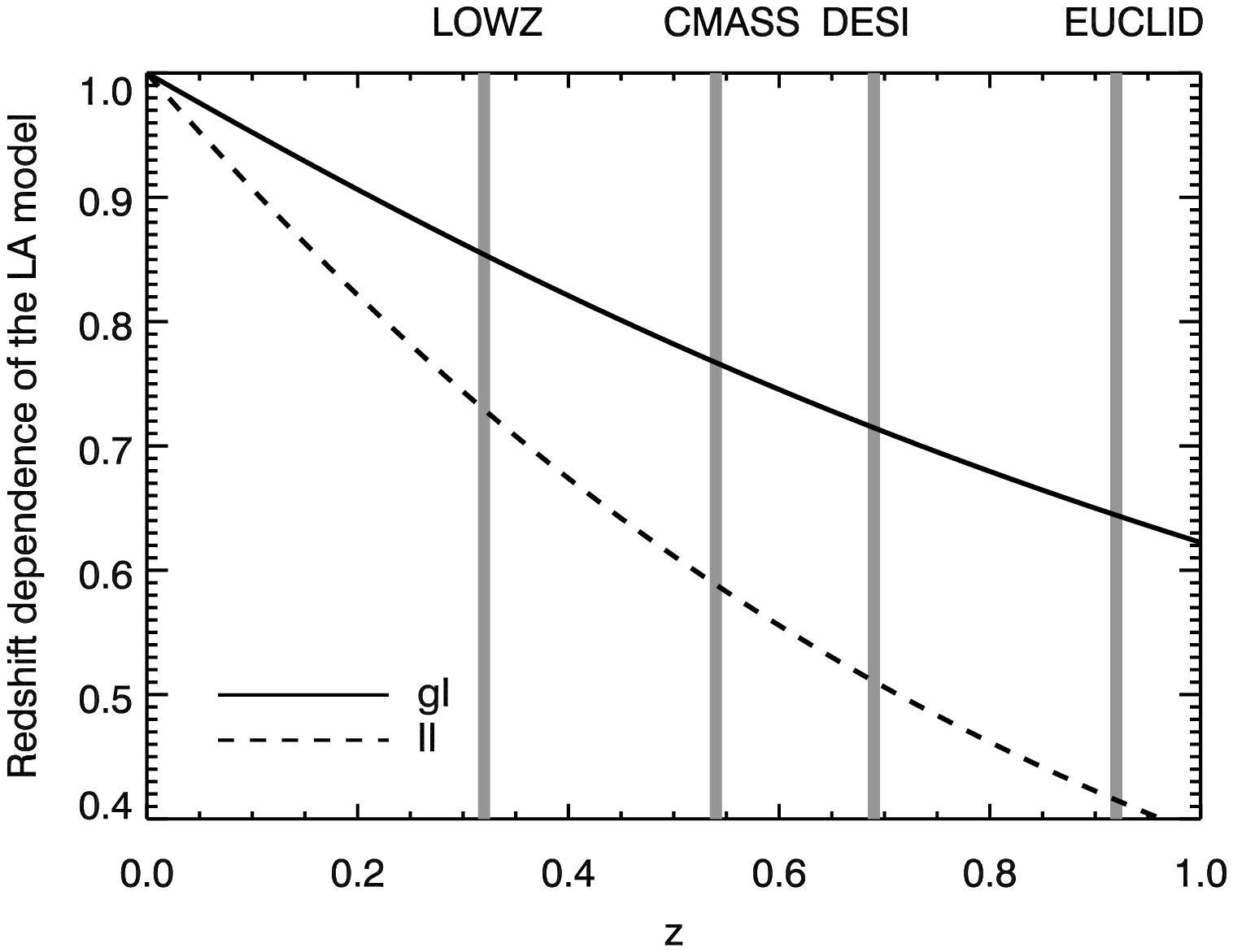}
\caption{The ratio between the redshift evolution of the primordial alignment model and the instantaneous model. Vertical lines in gray indicate the median redshift of the surveys considered in this work, from low to high redshift: LOWZ and CMASS (BOSS), DESI and EUCLID.}
\label{fig:redshiftdep}
\enf

\subsection{Contaminants to intrinsic alignments}
\label{sec:contam}

In Eq. \eqref{eq:corrobs}, we have assumed that the only contribution to the cross-correlation between the observed density field and the observed shapes comes from intrinsic alignments. There are additional contributions to the cross-correlation of the galaxy density field with galaxy shapes. The shear of a background galaxy by a lens gives a contribution that is referred to as galaxy-galaxy lensing ($gG$, the topic of Section \ref{sec:galaxy-galaxy}). At low redshift, there is a short baseline for matter fluctuations along the line of sight, hence the $gG$ contamination can be neglected \cite{Blazek11}. At higher redshifts, the degree of contamination depends on the accuracy of galaxy redshifts. For intrinsic alignments, the correlation $gI$ arises between galaxies that are physically associated and hence, separated by small physical distances. For galaxy-galaxy lensing, $gG$, the contribution to the correlation peaks for pairs of galaxies such that the background source is at twice the angular diameter distance between the lens and the observer. When redshifts are photometric, there is significant scatter of galaxies in redshift space such that a galaxy that in reality is behind a lens can be observed to be in the foreground, mixing the $gI$ and $gG$ signals.

In the most pessimistic scenario, where only photometric redshifts would be available for our fiducial targets, the increased uncertainty in the redshifts of the galaxies will induce a higher contamination of galaxy-galaxy lensing to the $gI$ correlation measurement that can readily be modeled, as done by \cite{Joachimi:2010xb} in the context of the MegaZ-LRG sample. In this context, a joint analysis of alignments, clustering and weak lensing cosmological constraints as in \cite{Joachimi10} would be more adequate. Nevertheless, while we consider only spectroscopic redshifts in this work, typical photometric scatter for LRGs can be significantly smaller than for galaxies selected for cosmic shear \cite{Benitez09}. Moreover, since LRGs have been shown to trace group environments \cite{Zheng09}, it is also conceivable that a combination of photometric redshifts for these galaxies and the group members can result in redshift estimates for our fiducial targets that would be more accurate than using photometric redshifts alone. 

The density field is also subject to lensing bias \cite[][]{Schmidt09}: magnification increases the density of galaxies at a given redshift by increasing the brightness of galaxies in the background. Lensing bias contributes a term $q\kappa$ to the weighting of the tidal field by the observed galaxy population, where $\kappa$ is the convergence field \cite{Bartelmann01} and $q(z)$ is the faint end slope of the luminosity function of LRGs. This introduces additional terms $mI$ and $mG$ in Eq. \eqref{eq:corrobs}. The term $mG$ arises when a galaxy is being magnified by the same matter overdensity that is shearing the second galaxy, and $mI$ arises when a galaxy in the background is being magnified by the same matter overdensity that is producing the tidal alignment of the second galaxy. 

We will only consider spectroscopic redshifts in this work. For the four samples of LRGs considered, we place an upper limit to the contamination of $gG$, $mG$ and $mI$ to the intrinsic alignment correlation. We apply the expressions derived by \cite{Joachimi:2010xb} in their Section $4.2$ for the angular power spectra of each cross-correlation at the median redshift of each sample. We limit the integral to small line of sight separations (i.e., within $[-\Pi_{\rm max},\Pi_{\rm max}]$) and assume that there is no scatter in the redshifts of the tracers. We estimate the slope of the luminosity function required for computing the magnification correlations using the expression derived in the Appendix of \cite{Joachimi10} and we assume limiting magnitudes of $r_{\rm lim}=\{22.5,23.5,24.5\}$ for the Baryon Oscillation Spectroscopic Survey\footnote{http://www.sdss3.org/surveys/boss.php} (BOSS) \cite{Dawson13}, the Dark Energy Spectroscopic Instrument (DESI) \cite{DESIwhite} and the EUCLID mission \cite[][]{EUCLIDRed}, respectively. Under these assumptions, we find that the $gG$ signal contributes less than $10\%$ to $gI$ for all surveys. The contribution of $mG$ is at least two orders of magnitude below the $gI$ correlation and while $mI$ is at the level of a few percent for LOWZ (LRGs at $0.2<z<0.4$ in BOSS), it is several orders of magnitude below $gI$ for CMASS (described in more detail in Section \ref{sec:cosmo}), DESI and EUCLID. The contaminations would decrease further if we decrease the line of sight correlation length, $\Pi_{\rm max}$. While we have considered $\Pi_{\rm max}=80 $Mpc$/h$ to ease the comparison of our results to those of \cite{Blazek11}, it would be possible to reduce this value to $\Pi_{\rm max}=60 $Mpc$/h$, since there is little contribution to the $gI$ signal from above those scales \cite{Mandelbaum06,Hirata07}.

\section{Cosmology from intrinsic alignments}
\label{sec:cosmo}

In this section, we explore the cosmological information in the intrinsic alignment of LRGs. We consider three spectroscopic
surveys: BOSS, DESI and the EUCLID mission. For each of these experiments, we constrain their ability to measure primordial non-Gaussianity and the BAO signature from the projected cross-correlation function of the galaxy density field and their intrinsic shears. 

BOSS \cite{Dawson13} is an undergoing survey of the SDSS-III collaboration. Among its targets, it has gathered spectra for low redshift LRGs, the LOWZ sample \cite{Parejko13}, and it has extended the LRG target selection of SDSS-II to $z=0.7$ \cite{Eisenstein11,White11}, targeting bluer but massive galaxies in the CMASS galaxy sample, with a typical bias similar to that of the LRG population. The LOWZ sample has three times the typical comoving number density of SDSS DR6 LRGs in the redshift range of $0.2<z<0.4$, with the same selection cuts but extended to fainter objects. While $w_{g+}$ has not yet been measured for these samples, the linear alignment model has been shown to reproduce the alignments of the MegaZ-LRG sample \cite{Joachimi10}, with photometric redshifts up to $z=0.7$. Hence, we consider both LOWZ and CMASS galaxies as tracers of the intrinsic alignment signal. 

For the LOWZ sample, we consider a constant comoving number density of $3\times10^{-4}h^3$Mpc$^{-3}$ in the range of $0.2<z<0.4$.
The median redshift of the LOWZ sample is $\bar{z}=0.32$. We construct the redshift distribution of CMASS galaxies from spectroscopic redshifts obtained in SDSS-III data release 10 (DR10). Figure \ref{fig:comovn} shows the comoving number density as a function of redshift for the galaxies in LOWZ and CMASS. The median redshift of CMASS galaxies is $\bar{z}=0.54$. The area of the survey footprint, in the DR10 release, is $6,373$ deg$^2$. We do not attempt to model in detail the selection effects necessary for constructing a large-scale structure catalogue based on the CMASS sample as done in \cite{Anderson13}, we only expect to reasonably reproduce the comoving number density of CMASS galaxies as a function of redshift. In the final data release (DR12), BOSS is expected to have covered $10,000$ deg$^2$. In the following sections, we make forecasts both for the currently available sample of BOSS galaxies in DR10 and for DR12. For the LOWZ and CMASS samples, we assume the galaxies have a constant bias $b\simeq 2$ \cite{White11,Nuza13,Parejko13}.

DESI is a spectroscopic survey scheduled for operation between the years $2018$ and $2022$. It will cover $14,000$ deg$^2$ and it will target spectroscopic LRGs in the redshift range $0.1<z<1.1$. We consider the comoving number density of LRGs observed by DESI, shown in Figure \ref{fig:comovn}, as quoted in Table 3 of \cite{DESIwhite}. The median redshift of the LRG sample observed by DESI is $\bar{z}=0.7$. While DESI is a purely spectroscopic survey, we assume imaging will be available from a combination of other surveys.

The EUCLID mission \cite[][]{EUCLIDRed}, scheduled to launch in 2020, will perform a weak gravitational lensing survey of $2\pi$ sr ($20,000$ deg$^2$) combined with a spectroscopic survey for measuring BAO. We model the redshift distribution of LRGs in an EUCLID-type survey in the range $0.5<z<1.5$ with:

\beeq
\frac{dN}{dzd\Omega}(z) = z^a \exp\left[-\left(\frac{z}{z_0}\right)^b\right],
\label{eq:lrgsel}
\eneq

\noindent where $a=2$, $b=1.5$ and $z_0=0.64$  \cite{Smail94,Joachimi10}, with a resulting median redshift of $\bar{z}\sim 0.9$. We have chosen to normalize the distribution of Eq. (\ref{eq:lrgsel}) to approximately match the comoving number density of LRGs at lower redshift \cite{Zehavi05}. To assess the impact of the normalization choice in our results, we consider two normalizations: $n_0 \equiv n(z=0.32) = 3\times10^{-4}h^3$Mpc$^3$ and $n_0 = 4\times10^{-4}h^3$Mpc$^3$. We show the resulting comoving number density of EUCLID LRGs as a function of redshift in Figure \ref{fig:comovn}. Since the proposed spectroscopic targets in EUCLID are H-$\alpha$ emitters \cite{EUCLIDRed} (i.e., blue galaxies), LRG redshifts will be better constrained from photometry at a typical precision that will be $\sigma_z \sim 0.03(1+z)$ for the overall population of galaxies \cite{EUCLIDRed}, which we do not include in our current modelling. For $n_0 = 3\times10^{-4}h^3$Mpc$^3$ and $n_0 = 4\times10^{-4}h^3$Mpc$^3$, the number of LRGs in the EUCLID sample will be $6.5$ million and $8.7$ million, respectively. 

Both for DESI and EUCLID, we consider that the bias of these tracers changes with cosmological epoch, becoming larger at higher redshifts. As a toy model for this process, we fix the correlation length of galaxies by making the bias proportional to the growth function. To normalize the bias, we match the LRG bias of the sample of \cite{Okumura09} at $\bar{z}=0.32$. This implies an almost linear increase in the bias from $b\sim 2.3$ to $b\sim 3.6$ in the range $0.5<z<1.5$.

\bef
\centering
\includegraphics[width=0.5\textwidth]{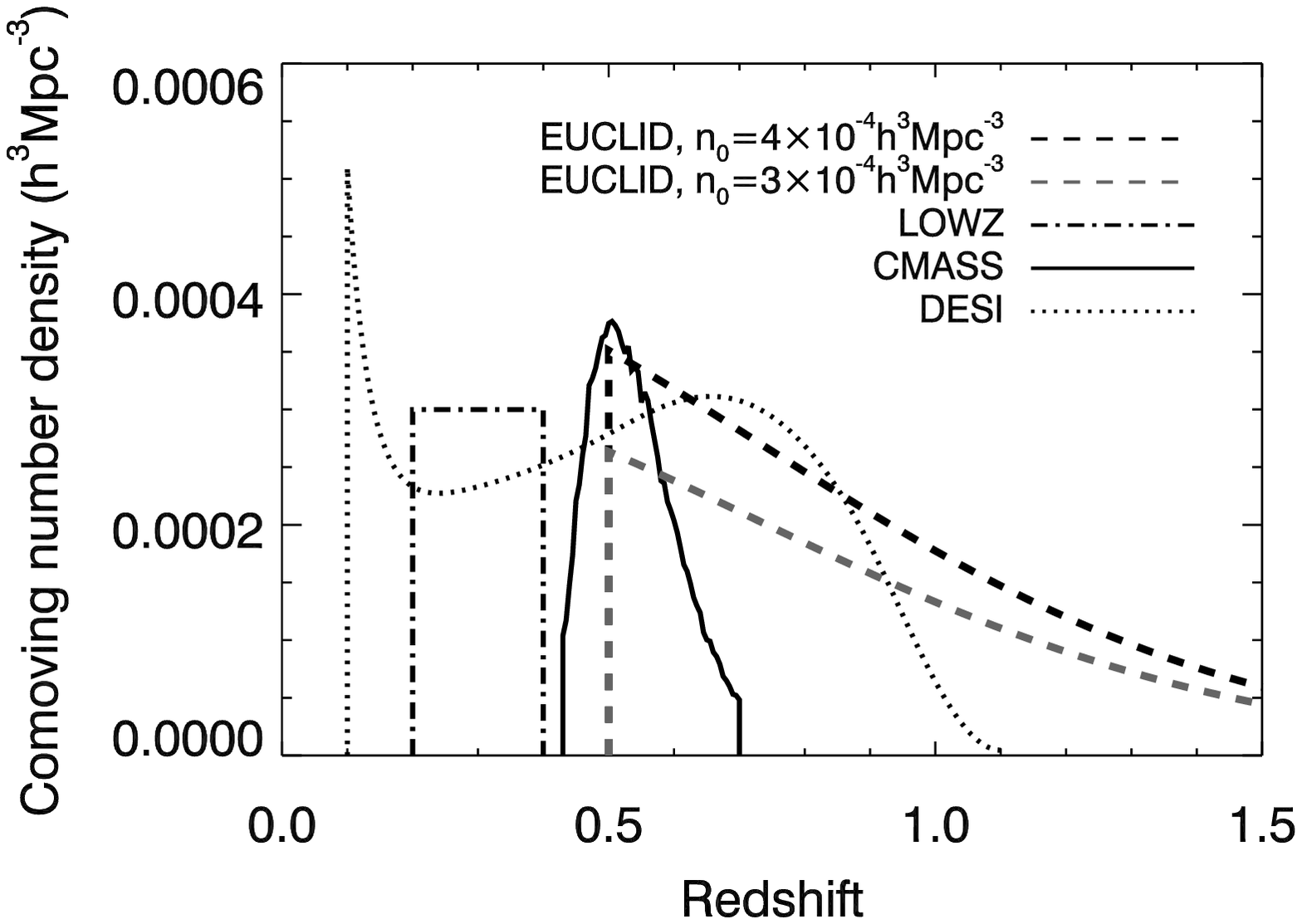}
\caption{The comoving number density of LOWZ (dashed-triple dotted line), CMASS galaxies as observed in DR10 (solid line), DESI LRGs (dashed) and EUCLID LRGs (dotted), as a function of redshift. See Section \ref{sec:cosmo} for a discussion of the underlying assumptions.}
\label{fig:comovn}
\enf

The typical systematic errors on galaxy shapes are subdominant compared to shape noise, and we consider a dispersion in the distortion of $\sigma_\gamma=0.25$ per component for all scenarios. 

\subsection{Primordial non-Gaussianity}
\label{sec:fnl}

We consider a simple model of local non-Gaussianity (NG) where the non-Gaussian primordial gravitational potential is given by

\beeq
\phi_{\rm NG} = \phi_p + \fnl \phi_p^2,
\eneq

\noindent where $\phi_p$ is the primordial Gaussian potential and $\fnl$ is a constant that parametrizes deviations from Gaussianity \cite{Komatsu01}. A non-zero detection of $\fnl$ would rule out any model of single-field inflation \cite{Creminelli:2004yq}. Current best constraints on the local type $\fnl$ have been set by the {\it Planck} Collaboration at $\fnl=2.7\pm5.8$ ($68\%$ C.L.) from observations of the Cosmic Microwave Background temperature field \cite{PlanckNG}. In the near future, tighter constraints will come from large-scale structure surveys, with EUCLID being capable of obtaining constraints to $\Delta \fnl=2$ from galaxy clustering \cite[][]{EUCLIDRed}.

The non-Gaussian density field can be derived from the Poisson equation to first order in $\fnl$ as

\beeq
\delta^{\rm NG}\approx\delta^{\rm G}\left[1+2\fnl\phi_p\right],
\label{eqn:dng}
\eneq

where $\delta^{\rm G}$ is the Gaussian density field. We can see from Eqs. \eqref{eqn:ermsorig} and \eqref{eqn:dng} that primordial non-Gaussianity will have an effect on the ellipticity of the galaxies. The observable $\langle e^2 \rangle$ behaves similarly to $\sigma_R^2$, the rms variation of mass within a Gaussian sphere of radius $R$. In the presence of primordial non-Gaussianity, there will be additional terms in Eq. \eqref{eqn:ermsorig}. We can separate the contribution to the gravitational potential from long, $\phi_l$, and short wavelength modes, $\phi_s$, following the peak-background split formalism \cite{Cole89,Slosar08}, 

\beeq
\phi_p = \phi_l+\phi_s,
\eneq

\noindent and in doing so, we find that the effect of primordial non-Gaussianity on the scatter of intrinsic ellipticities is 

\beeq
\langle e^2 \rangle \rightarrow \langle e^2 \rangle (1+4 f_{\rm NL}\phi_l).
\eneq

On the other hand, local non-Gaussianity gives rise to a scale dependent bias of halos on large scales \cite[][]{Dalal08, Slosar08}:

\beeq
\Delta b (k,z) = 3 \fnl (b-p) \frac{\delta_c\Omega_M}{k^2 T(k) D(z)} \frac{H_0^2}{c^2},
\eneq

\noindent where $c$ is the speed of light, $T(k)$ is the linear matter transfer function at $z=0$, $\delta_c=1.686$ is the spherical collapse linear overdensity and we have assumed a merger history consistent with $p=1$, implying that LRGs have not undergone recent mergers. This assumption is consistent with the assumption of passive evolution of LRGs in the LA model, where the halo shape is determined at $z_p$.

The components of the intrinsic shear field in Fourier space in the non-Gaussian case are

\beeq
\gamma^{I,{\rm NG}}(\kbf,z) = \iacons\frac{(k_x^2-k_y^2,2k_xk_y)}{k^2}\delta^{\rm NG} (\kbf,z),
\eneq

\noindent where the non-Gaussian density field in Fourier space is given by

\beeq
\delta^{\rm NG} (\kbf,z) = \delta^{\rm G} (\kbf,z) + \fnl \frac{3H_0^2\Omega_M(1+z)}{2 c^2D(z)} k^2 \int \frac{d^3\kbf_1}{(2\pi)^3} \frac{\delta^{\rm G} (\kbf-\kbf_1,z)\delta^{\rm G} (\kbf_1,z)}{|\kbf-\kbf_1|^2|\kbf_1|^2},
\eneq

\noindent and the observed galaxy density field is

\beeq
\delta_g^{\rm NG}(\kbf,z) = [b+\Delta b (k,z)] \delta^{\rm G} (\kbf,z).
\eneq

An additional effect of primordial non-Gaussianity is to modify the effect of RSD on large scales by changing the bias (i.e., replacing $b$ by $b+\Delta b$ in the expression for $\beta$ in Eq. (\ref{eq:rsdsm})) and to add terms to the Kaiser factor \cite{Schmidt10}. While the change in the bias produces a significant effect that we take into account in this work, additional terms become relevant only on non-linear scales, smaller than $k\gtrsim 0.1h/$Mpc \cite{Schmidt10}. On small scales, the non-linearity of the density field can also give rise to non-Gaussian effects on the intrinsic alignments of galaxies. \cite{Hui08} have explored these effects for a different alignment model than the one considered in this work. However, in our case, because we restrict to large scales, this effect can be safely neglected. 

\subsubsection{Intrinsic ellipticity-density cross-correlation}

In Section \ref{sec:observed}, we introduced the correlation functions of the galaxy density field and the observed shears. 
The full non-Gaussian contribution to the observed density-intrinsic shear correlation in Fourier space is given by 

\bear
\langle \delta_g^{\rm NG}(\kbf',z)\tilde{\gamma}^{*}(\kbf,z)\rangle &=& \langle [b+\Delta b (k',z)]\delta^{\rm G}(\kbf',z)[ \gamma^{I,{\rm NG}}(\kbf,z) \nonumber\\
&+& \int d^3\kbf_1 \frac{b+\Delta b(k_1,z)}{(2\pi)^3}\,\gamma^{I,{\rm NG}}(\kbf-\kbf_1,z)\delta^{\rm G}(\kbf_1,z)]^{*}\rangle,
\label{eq:corrobsNG}
\enar

\noindent To order $\fnl$, Eq. \eqref{eq:corrobsNG} can be simplified to

\bear
\langle \delta_g^{\rm NG}(\kbf',z)\tilde{\gamma}^{*}(\kbf,z)\rangle &=& -\iacons [b+\Delta b (k,z)](2\pi)^3\delta^{(3)}(\kbf-\kbf') \tidal\frac{P_s(\kbf,z)}{k^2} \nonumber\\
&-& 3b^2 \fnl \frac{H_0^2C_1\rho_{\rm crit}\Omega_M^2}{c^2D^2(z)} (1+z) \delta^{(3)}(\kbf-\kbf') \frac{P_s(\kbf,z)}{k^2}\nonumber\\
&\times& \int d^3\kbf_1 (k_{2x}^2-k_{2y}^2,2k_{2x}k_{2y}) \frac{P_s(\kbf_1,z)}{k_1^2},
\label{eq:mix}
\enar

\noindent where $\kbf_2=\kbf-\kbf_1$ and components $(k_{2x},k_{2y})$ on the plane of the sky.

The non-Gaussian correlation function, integrated over $z$, is 

\bear
w_{g+}^{NG}(r_p) &=& \int dz \mathcal{W}(z)  \frac{1}{\pi^2} {C_1\rho_{\rm crit}\Omega_M\over D(z)} \int_0^\infty dk_z \int_0^\infty dk_{\perp} \frac{k_{\perp}^3}{(k_{\perp}^2+k_z^2)k_z}P_s(\kbf,z)\sin(k_z\Pi_{\rm max})J_2(k_{\perp} r_p)\nonumber\\
&& \times\left[b+\Delta b(k,z)+\frac{3b^2\fnl H_0^2\Omega_M(1+z)}{2\pi^2D(z)}\int_{0}^\infty dk_{1z}\int_{0}^\infty dk_{1\perp} k_{1\perp} \frac{P_s(\kbf_1,z)}{k_1^2}\right].
\label{eq:ngwgp}
\enar

There are two non-Gaussian terms to order $\mathcal{O}(\fnl)$ contributing to $w_{g+}$. The second term inside the brackets in Eq. (\ref{eq:ngwgp}) is the usual non-Gaussian bias on large scales, derived by \cite{Dalal08}. The mixing of scales in the second term of Eq. (\ref{eq:mix}) gives rise to the third term inside the brackets. Interestingly, this term gives a constant constribution with scale. In all terms in Eq. (\ref{eq:ngwgp}), the bias that appears in the Kaiser factor of Eq. (\ref{eq:rsdsm}) is replaced by the non-Gaussian large-scale bias, $b+\Delta b$.

The relative contribution from the third term to $w_{g+}$ is at the level of $10^{-5}$. Moreover, because this term adds a constant to the bias, its effect is to rescale the correlation function. Given that the strength of the alignment cannot be derived from first principles, in practice it is not feasible to distinguish between this constant increase in the bias and intrinsic evolution of $C_1$ as a function of redshift at the level of $0.001\%$.

\bef
\centering
\subfigure[\hskip 2pt LOWZ. DR12 points have been artificially displaced to higher $r_p$ for visualization purposes.]{
\includegraphics[width=0.45\textwidth]{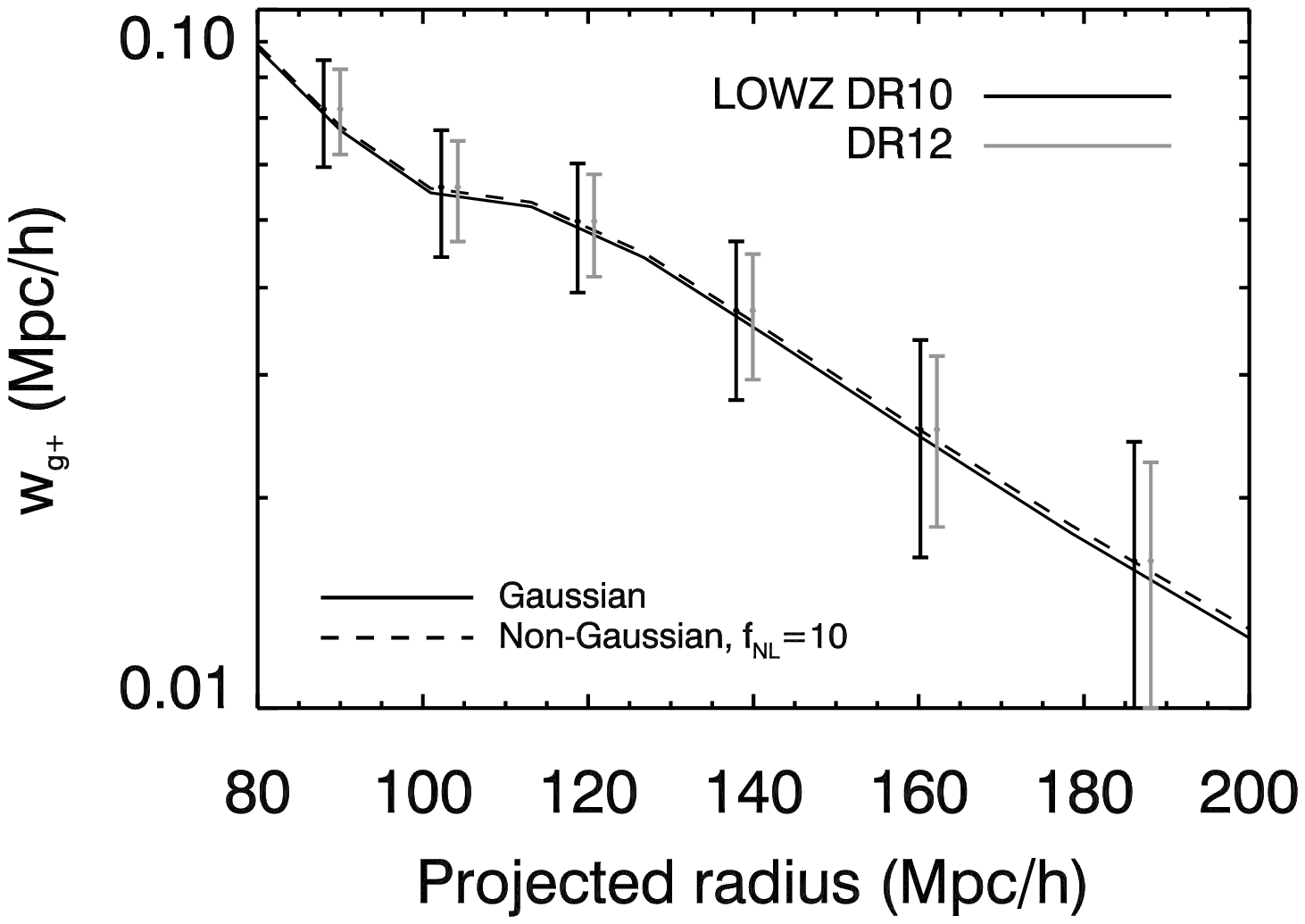}
}
\subfigure[\hskip 2pt CMASS. DR12 points have been artificially displaced to higher $r_p$ for visualization purposes.]{
\includegraphics[width=0.45\textwidth]{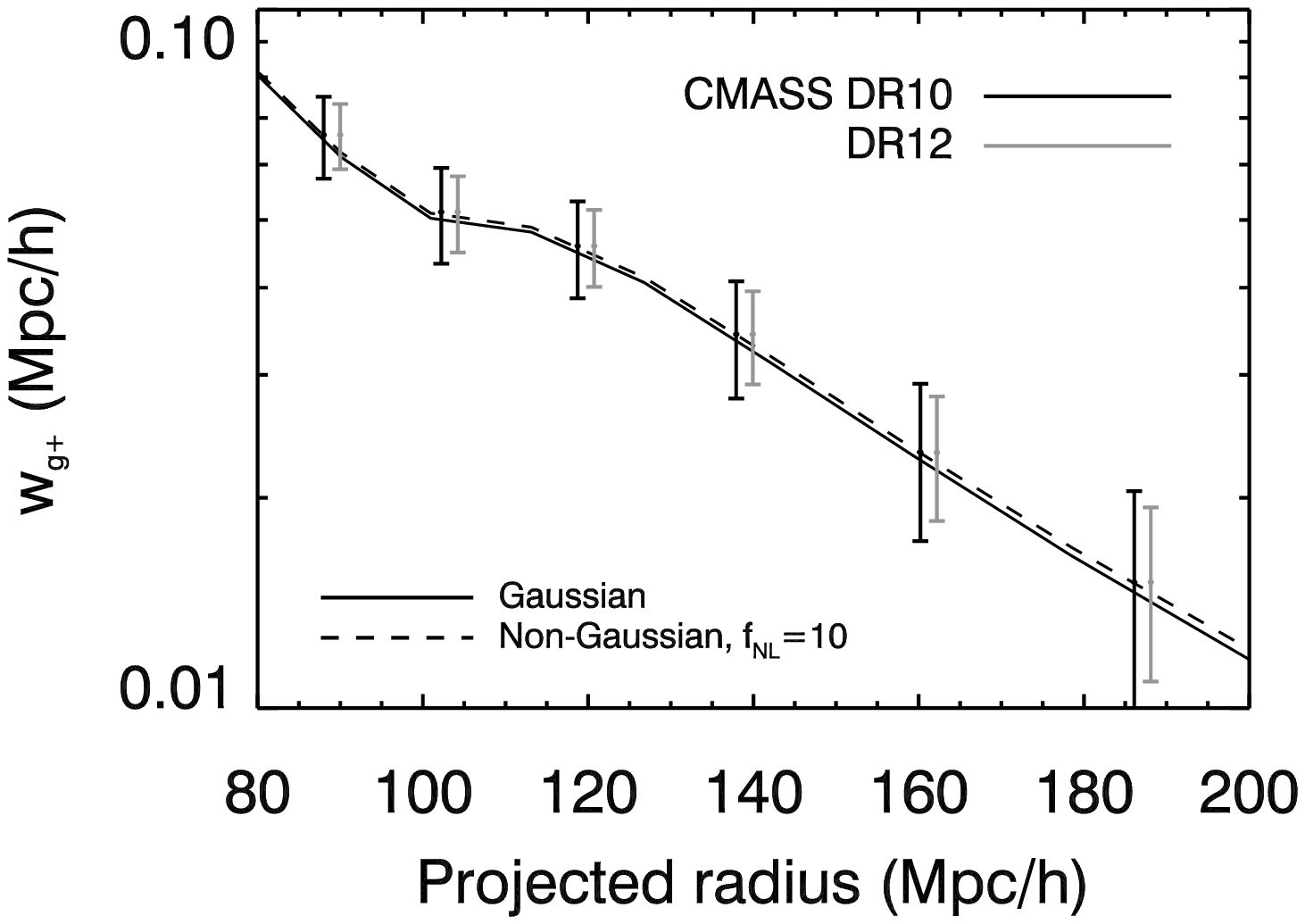}
}
\subfigure[\hskip 2pt DESI.]{
\includegraphics[width=0.45\textwidth]{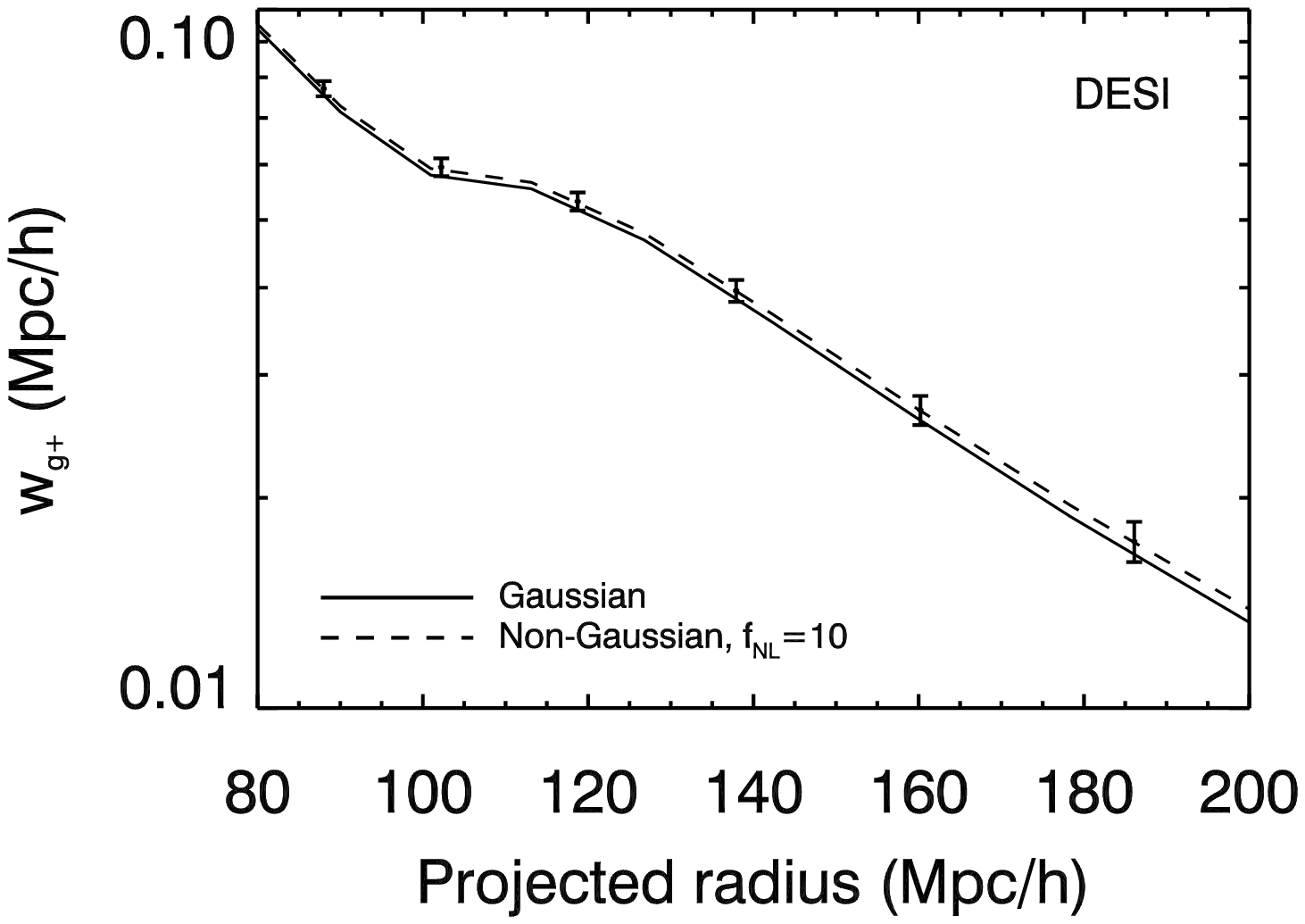}
}
\subfigure[\hskip 2pt EUCLID.]{
\includegraphics[width=0.45\textwidth]{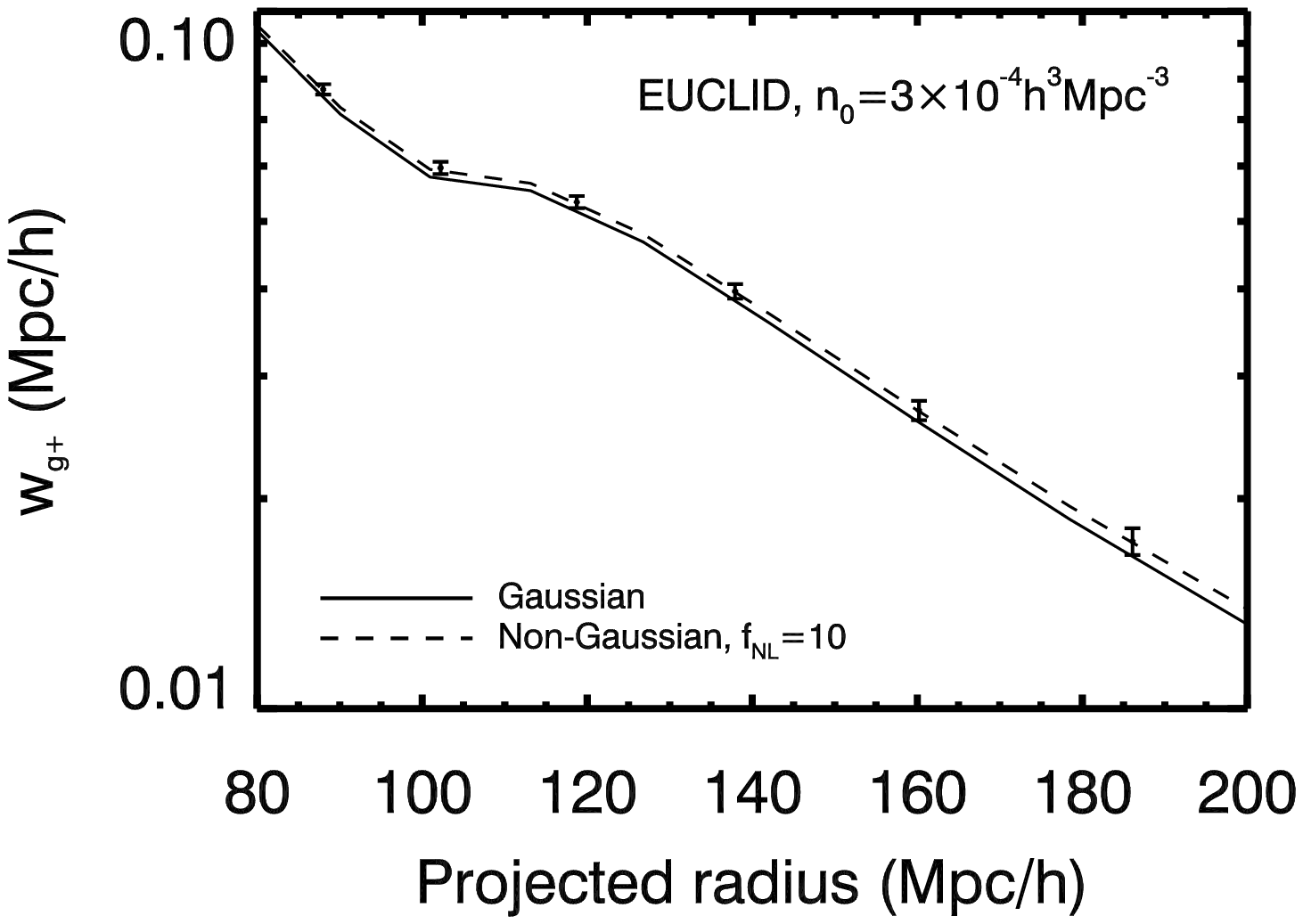}
}
\caption{The correlation function of galaxies and $+$ ellipticities in a Gaussian cosmology
(solid line) and a cosmology with primordial non-Gaussianity of $\fnl=10$ (dashed line), with a smoothing scale of $k_{\rm max}=10$ h$/$Mpc
for the different surveys considered (LOWZ and CMASS in DR10 and DR12, DESI and EUCLID).}
\label{fig:correlfnl}
\enf

In Figure \ref{fig:correlfnl} we show the cross-correlation function $w_{g+}$ for a cosmology with $f_{NL}=10$ compared to a Gaussian correlation function for the four samples of LRGs described in Section \ref{sec:cosmo}. The error bars have been computed by constructing the covariance matrix of the cross-correlation function, as described in the Appendix of this work. The correlation function has been tested for convergence due to numerical integration and our tests indicate un upper limit in the uncertainty due to convergence of $0.25\%$.

The likelihood difference between a non-Gaussian and a Gaussian $w_{g+}$ is obtained by summing over $(i,j)$ projected radius bins

\beeq
\Delta\chi^2=\sum_{i,j}\left(w_{g+,i}^{\rm NG}-w_{g+,i}\right)C_{i,j}^{-1}\left(w_{g+,j}^{\rm NG}-w_{g+,j}\right),
\eneq

\noindent where $C_{i,j}$ is the covariance matrix. Table \ref{table:fnl} shows the estimated signal-to-noise ratios for detecting non-Gaussianity of $f_{NL}=10$ for each sample up to $200$ Mpc$/h$. For DESI and EUCLID, extending the constraints to $r_p<500$ Mpc$/h$ does not alter our predictions significantly. To show the impact of cosmic variance on these results, we include in Table \ref{table:fnl} the constraints on the detection of non-Gaussianity in the case where cosmic variance is neglected. While shape noise typically dominates the covariance matrix on small scales, cosmic variance increases with scale, as does the effect of non-Gaussianity. For all samples considered, the $S/N$ for detecting $f_{NL}=10$ is $<2$ when cosmic variance is taken into account. As a consequence, for ongoing and upcoming surveys, primordial non-Gaussianity needs not be considered when removing the intrinsic alignments signal from gravitational lensing correlations.

\begin{table}[!h]
\small
\centering
\caption{Signal-to-noise ratio for the detection of primordial
non-Gaussianity of $\fnl=10$ in the cross-correlation of the galaxy density field
with intrinsic ellipticity for the surveys considered in this work. We have defined $n_0=\eta\times10^{-4}h^3$Mpc$^{-3}$.}
\label{table:fnl}
\vskip 3pt
\begin{tabular*}{0.552\textwidth}{| c || c  c | c  c | c | c  c |}
\hline
Survey & \multicolumn{2}{|c|}{LOWZ} & \multicolumn{2}{|c|}{CMASS} & $\,$DESI$\,$ & \multicolumn{2}{|c|}{EUCLID}  \\
       & DR10$\,$ & DR12$\,$ & DR10$\,$ & DR12$\,$ &  & $\eta=3\,\,\,$ & $\eta=4\,\,\,$ \\
\hline
With cosmic variance & \SNfnllowz & \SNfnllowzTWELVE &  \SNfnlboss  & \SNfnlbossTWELVE & \SNfnldesi & \SNfnlEUCLID  & \SNfnlEUCLIDfour\\
Without cosmic variance &  \SNfnllowzjs & \SNfnllowzjsTWELVE & \SNfnlbossjs & \SNfnlbossjsTWELVE & \SNfnldesijs & \SNfnlEUCLIDjs & \SNfnlEUCLIDjsfour \\
\hline
\end{tabular*}
\end{table}

In Eq. \eqref{eq:rsdsm}, we introduced the Kaiser factor, the correction factor to the matter power spectrum when RSD are taken into account. In Figure \ref{fig:rsd}, we show the impact on the correlation function when the Kaiser factor is included compared to when its effect is ignored. In the right panel of Figure \ref{fig:rsd}, we also show the change in the $w_{g+}$, with RSD, when we apply a non-Gaussianity of $\fnl=10$. On large scales, the effects of RSD and primordial non-Gaussianity are similar: they both produce an enhancement in the correlation function. This is due to the rapid oscillation of the integrand in Eq. \eqref{eq:wgp} along the length of the cylinder of $w_{g+}$. When projected, RSD and primordial non-Gaussianity have a similar scale-dependence, they both increase at large separations. Neglecting to model the effect of RSD can lead to a false detection of primordial non-Gaussianity of the local type. 

\bef
\centering
\subfigure[\hskip 2pt Relative difference between the Gaussian $w_{g+}$ with and without the effect of RSD.]{
\includegraphics[width=0.45\textwidth]{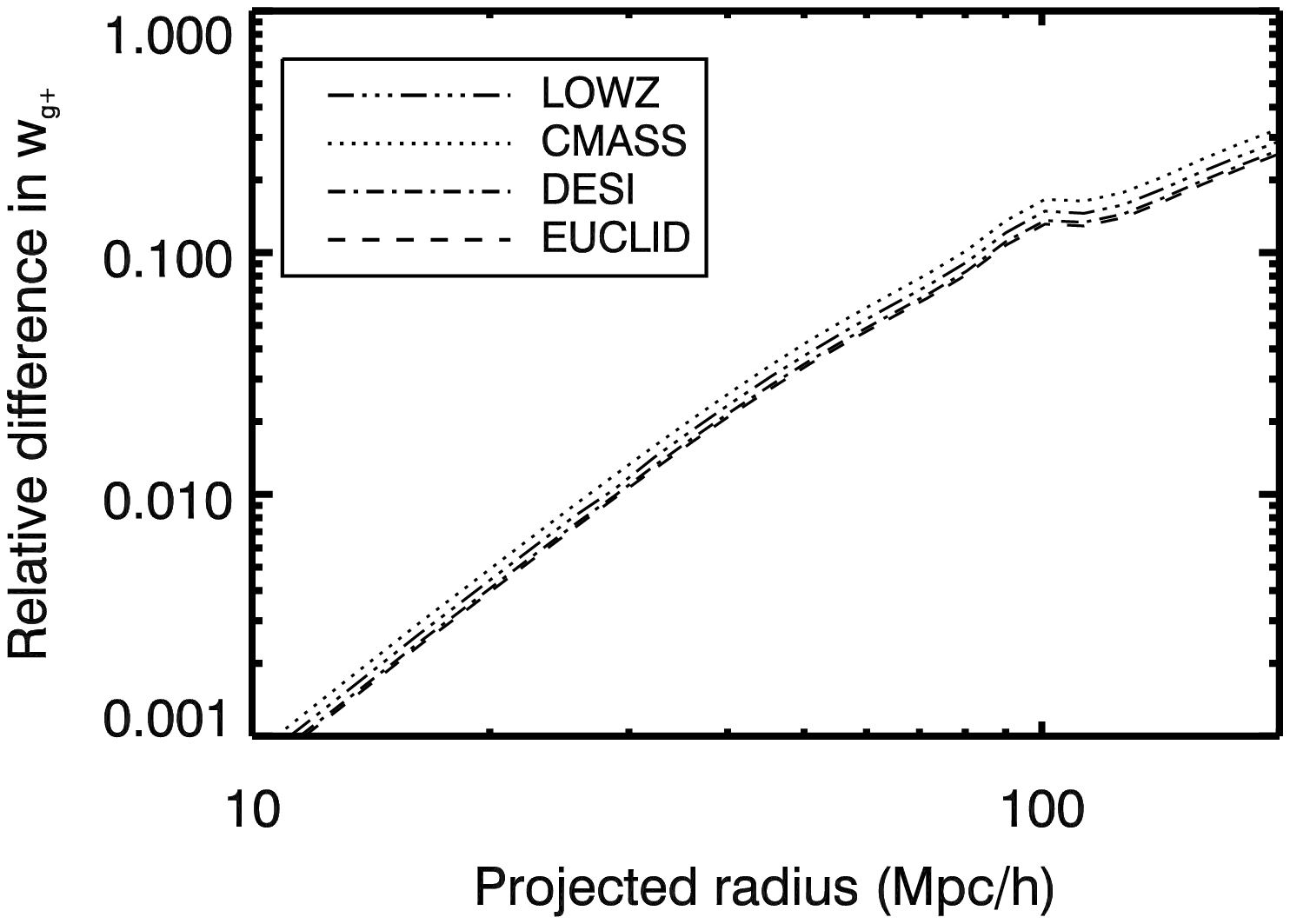}
\label{fig:rsd}
}
\subfigure[\hskip 2pt Relative difference between the Gaussian $w_{g+}$ and the non-Gaussian $w_{g+}$ for $\fnl=10$.]{
\includegraphics[width=0.45\textwidth]{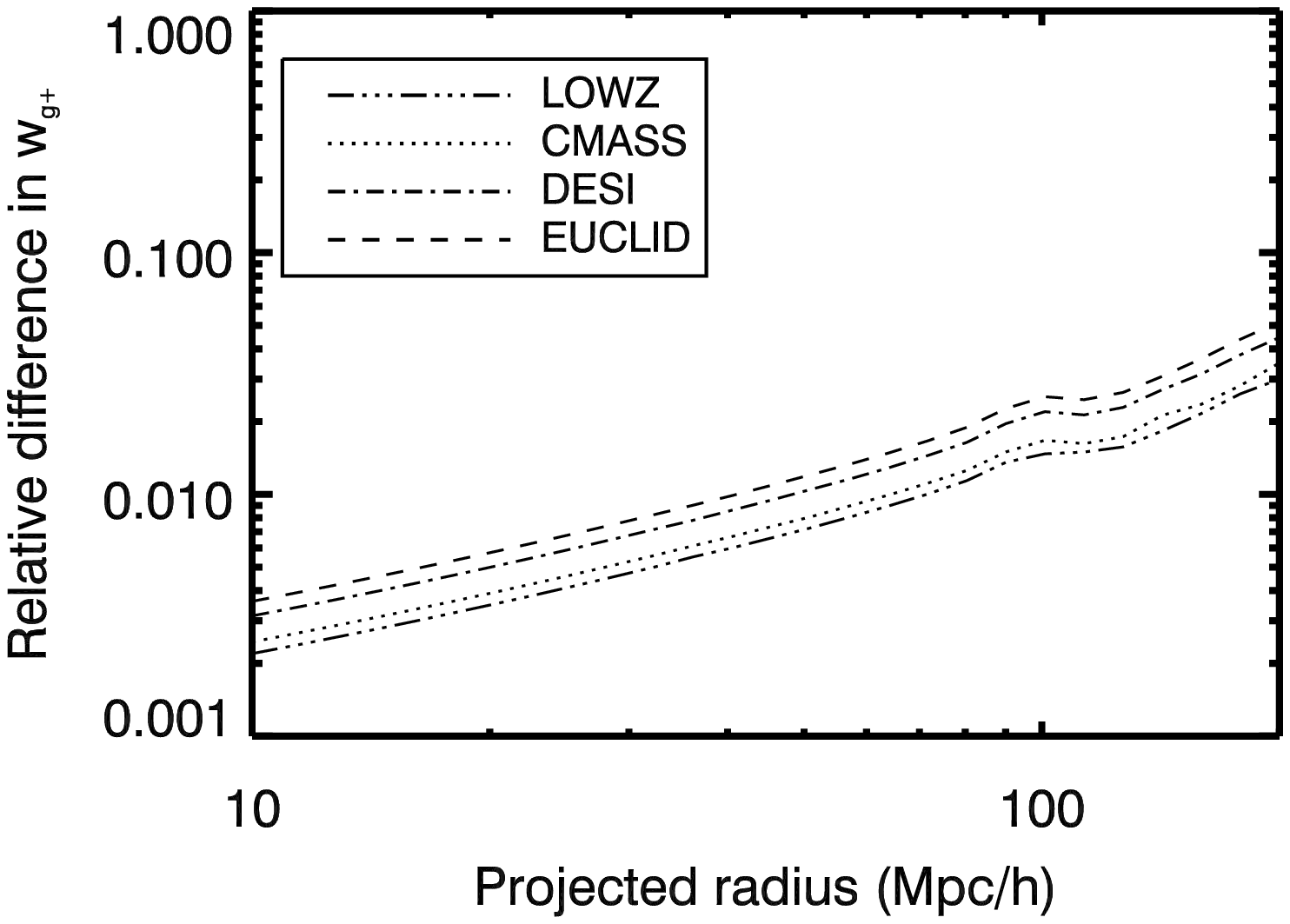}
\label{fig:ngrel}
}
\caption{Comparison of the effect of RSD on the cross-correlation function $w_{g+}$ with the effect of primordial non-Gaussianity of the local type parametrized by $\fnl=10$ for LOWZ, CMASS, DESI and EUCLID. In projection, the two effects have a similar scale-dependence, increasing at large separations.}
\enf

\subsection{Baryon Acoustic Oscillations}
\label{sec:bao}

We study the level of detectability of Baryon Acoustic Oscillations (BAO) in the intrinsic ellipticity-density field cross-correlation. We compute the likelihood difference between the case in which the cross-correlation has a matter power spectrum with baryon damping but no oscillations, $w_{g+}^{\rm no\,wiggles}$ \cite{Eisenstein:1997ik}, and the case in which the effect of the baryons is present, $w_{g+}^{\rm wiggles}$:

\beeq
\Delta\chi^2=\sum_{i,j}\left(w_{g+,i}^{\rm wiggles}-w_{g+,i}^{\rm no\,wiggles}\right)C_{i,j}^{-1}\left(w_{g+,j}^{\rm wiggles}-w_{g+,j}^{\rm no\,wiggles}\right).
\eneq

In Figures \ref{fig:baocorrel_LOWZ}-\ref{fig:baocorrel_EUCLID}, we show the two correlation functions with and without the BAO feature for BOSS, DESI and EUCLID. In these figures, we see clearly that the BAO appears in the $w_{g+}$ correlation successively as a trough, a node and a bump around the $110$ Mpc$/h$ scale. The usual bump observed in the matter correlation function appears at $\sim 100$Mpc$/h$, roughly coinciding with the location of the trough in $w_{g+}$. This feature has also been called a ``shoulder'' in the galaxy-galaxy lensing correlation function by \cite{Jeong09}. The correlation functions without wiggles have numerically converged to a tolerance of $<1\%$.

In Figure \ref{fig:baoSN}, we show the estimated cumulative signal-to-noise ratio of the detection of the BAO in the $w_{g+}$ cross-correlation when we take a fixed upper limit to the interval of $200$Mpc$/h$ and we vary the lower bound. The $S/N$ displays a significant increase at $\sim110$ Mpc$/h$. We do not extend our computation of the $S/N$ below the scale of $50$Mpc$/h$ due to differences in the prediction of $w_{g+}$ in the non-linear scales by CAMB/halofit and the analytical predictions of the case without BAO by \cite{Eisenstein:1997ik}. For the BOSS samples, the $S/N$ at $80$ Mpc$/h$ reaches $\SNbaolowz$ and $\SNbaocmass$ for LOWZ and CMASS in DR10, respectively, and $\SNbaolowzTWELVE$ and $\SNbaocmassTWELVE$ for those samples in DR12. For the LOWZ sample only, cosmic variance has a significant impact on the $S/N$ estimate due to the smaller volume probed at low redshift. For DESI, the cumulative $S/N$ predicted is $\SNbaodesi$ at $80$ Mpc$/h$ and for EUCLID, it is $\SNbaoEUCLID$ and $\SNbaoEUCLIDfour$ for with $n_0=3\times10^{-3}h^3$Mpc$^{-3}$ and with $n_0=4\times10^{-3}h^3$Mpc$^{-3}$, respectively. Thus, the BAO signature could be detected at high significance in DESI and EUCLID. In these surveys, large-scale structure probes will be combined to achieve tighter cosmological constraints. The addition of the $gI$ correlation to the ensemble of observables probed by these surveys is worth considering in the context of measuring the evolution of the distance scale of the BAO.

\bef
\centering
\subfigure[\hskip 2pt LOWZ. DR12 points have been artificially displaced to higher $r_p$ for visualization purposes.]{
\includegraphics[width=0.45\textwidth]{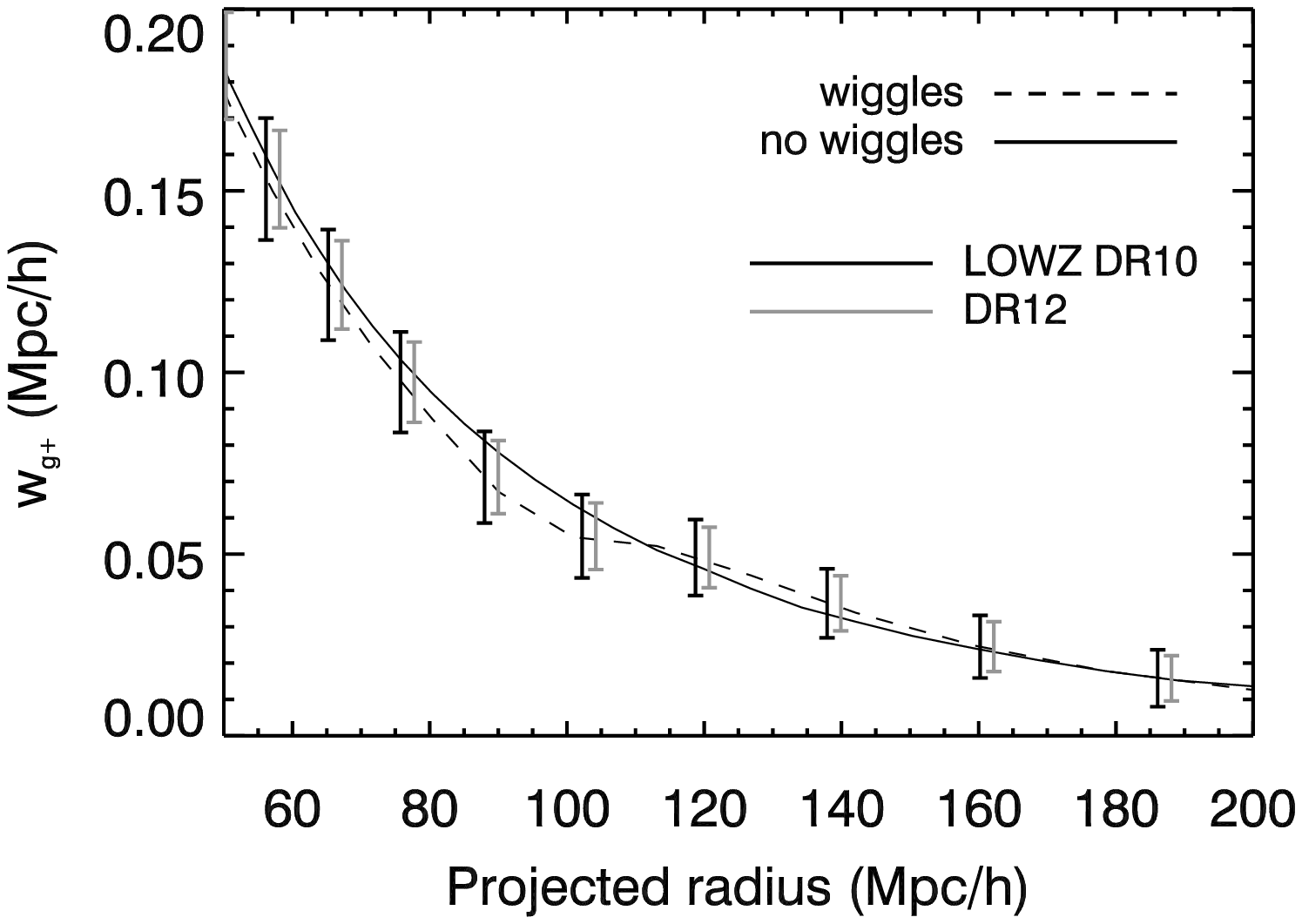}
\label{fig:baocorrel_LOWZ}
}
\subfigure[\hskip 2pt CMASS.DR12 points have been artificially displaced to higher $r_p$ for visualization purposes.]{
\includegraphics[width=0.45\textwidth]{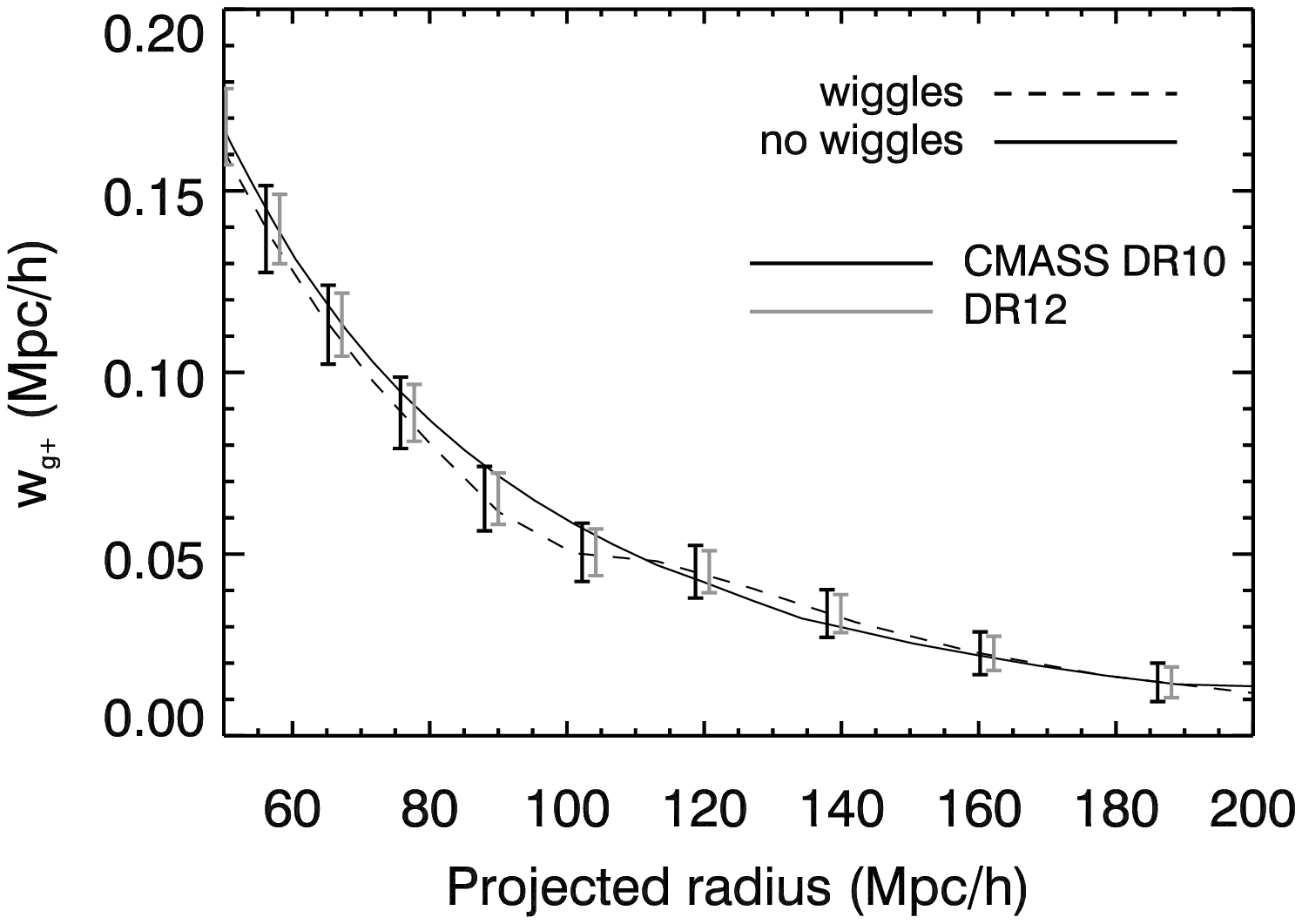}
\label{fig:baocorrel_CMASS}
}
\subfigure[\hskip 2pt DESI.]{
\includegraphics[width=0.45\textwidth]{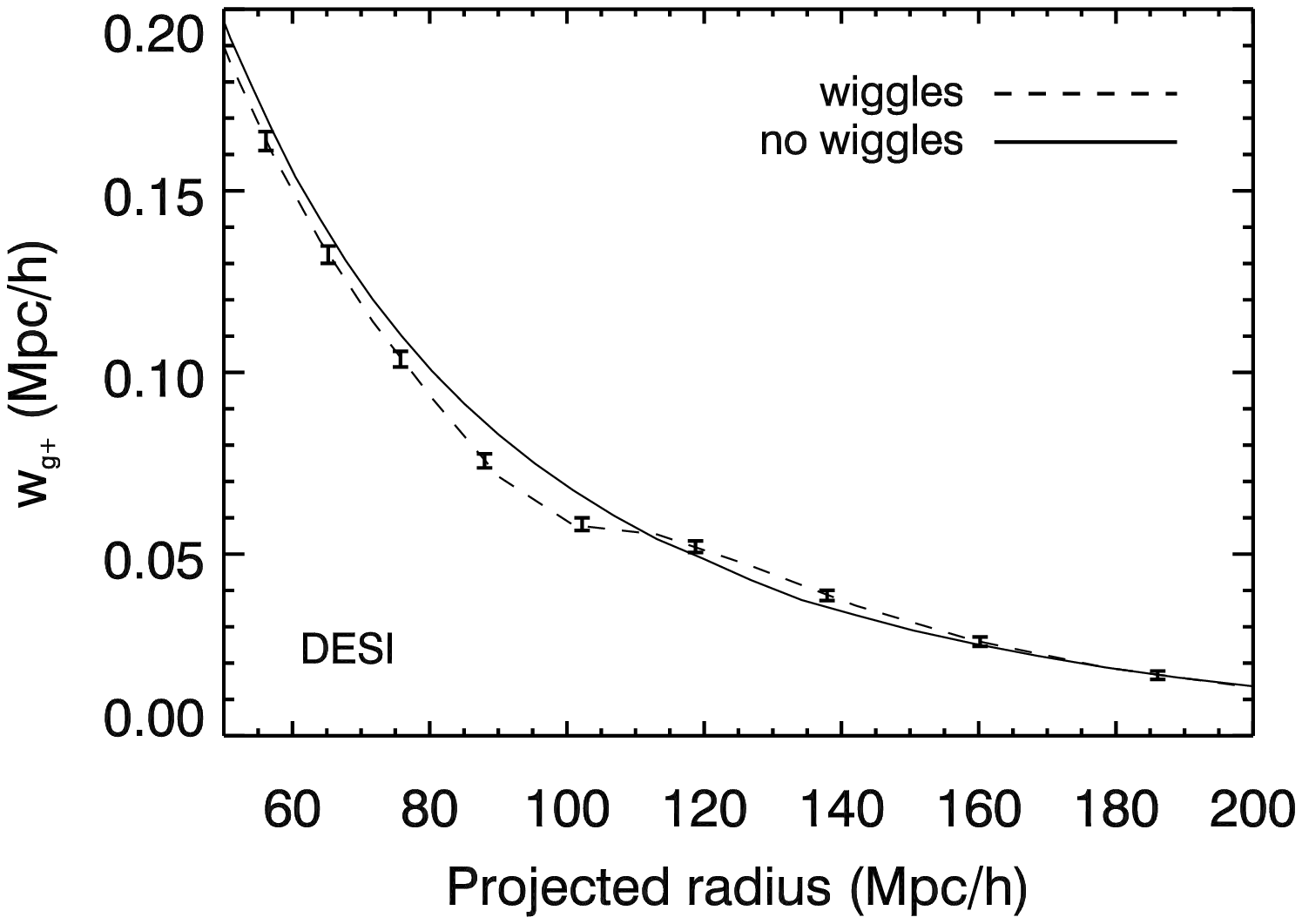}
\label{fig:baocorrel_DESI}
}
\subfigure[\hskip 2pt EUCLID ($n_0=3\times10^{-3}h^3$Mpc$^{-3}$).]{
\includegraphics[width=0.45\textwidth]{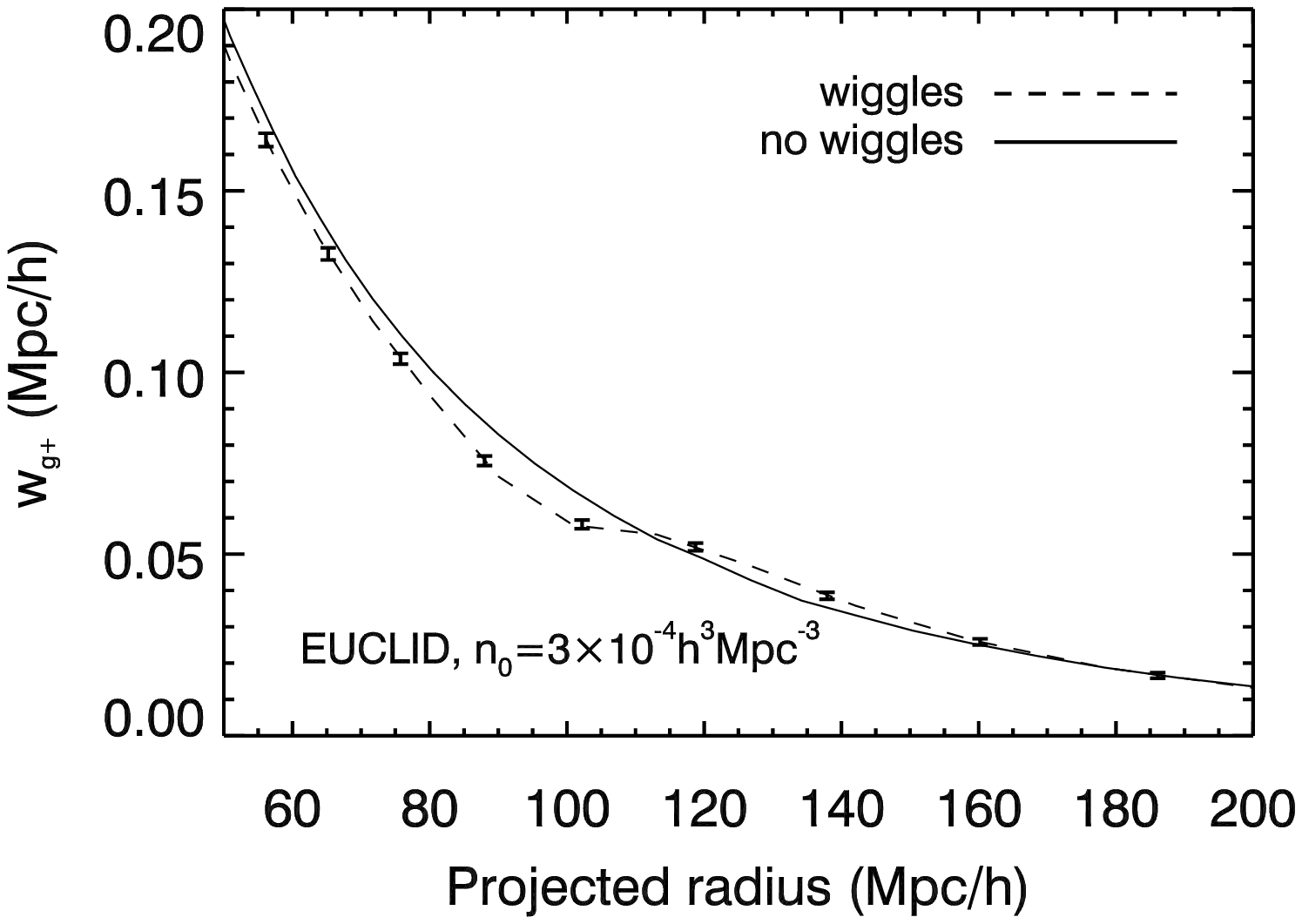}
\label{fig:baocorrel_EUCLID}
}
\caption{The correlation function of galaxies and $+$ ellipticities in the cases:
with baryons but no wiggles \cite{Eisenstein:1997ik} (solid line) and for our fiducial case (dashed line).
The data points correspond to the radially binned correlation function with errors predicted
from the covariance matrix. At $\sim 110$ Mpc$/h$, the BAO appears as a decrement, a node and an increment in $w_{g+}$ with respect to the ``no wiggles'' case, rather than the usual bump present in $w_{gg}$. }
\enf

\bef
\centering
\subfigure[\hskip 2pt LOWZ]{
\includegraphics[width=0.45\textwidth]{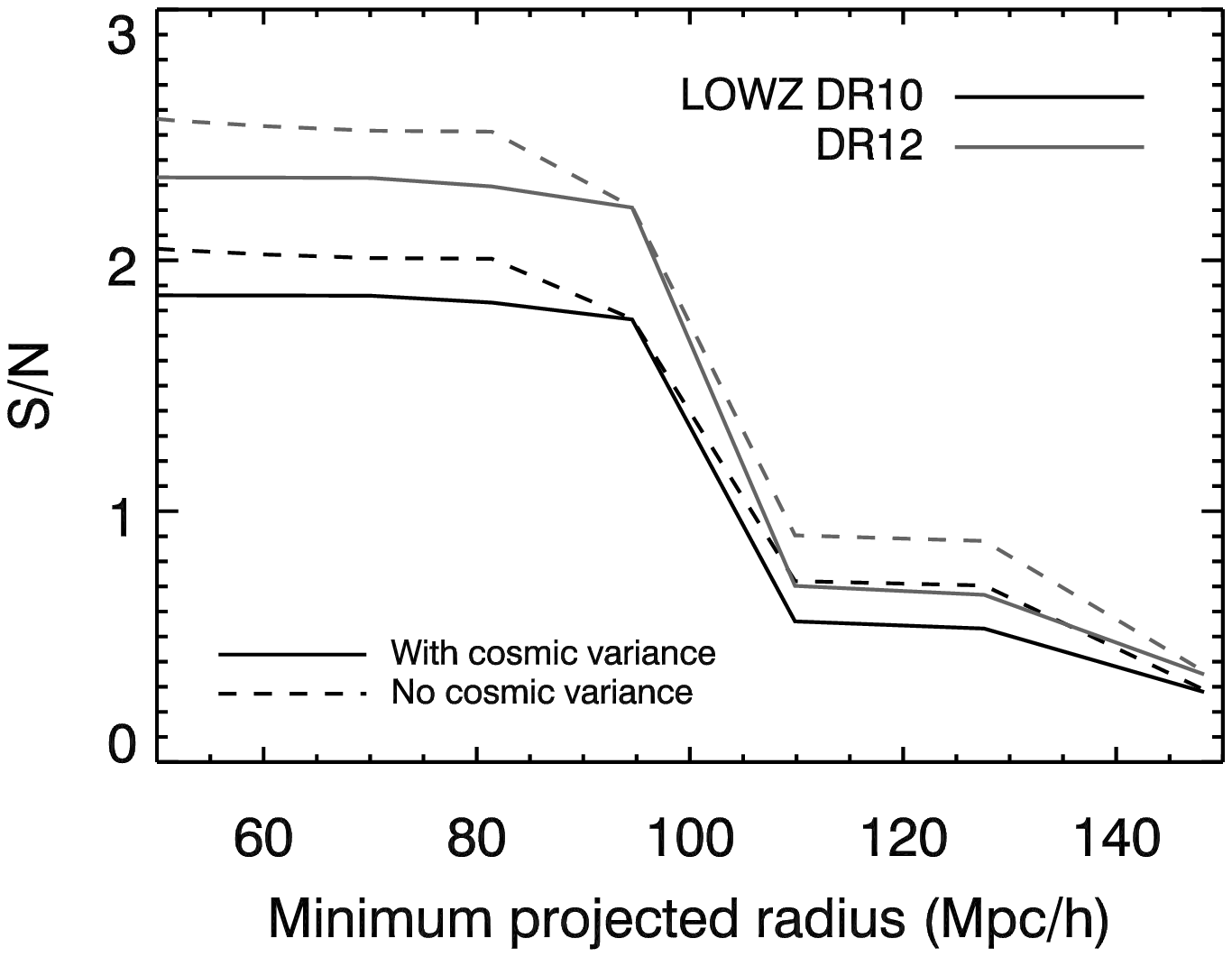}
\label{fig:snbaolowz}
}
\subfigure[\hskip 2pt CMASS]{
\includegraphics[width=0.45\textwidth]{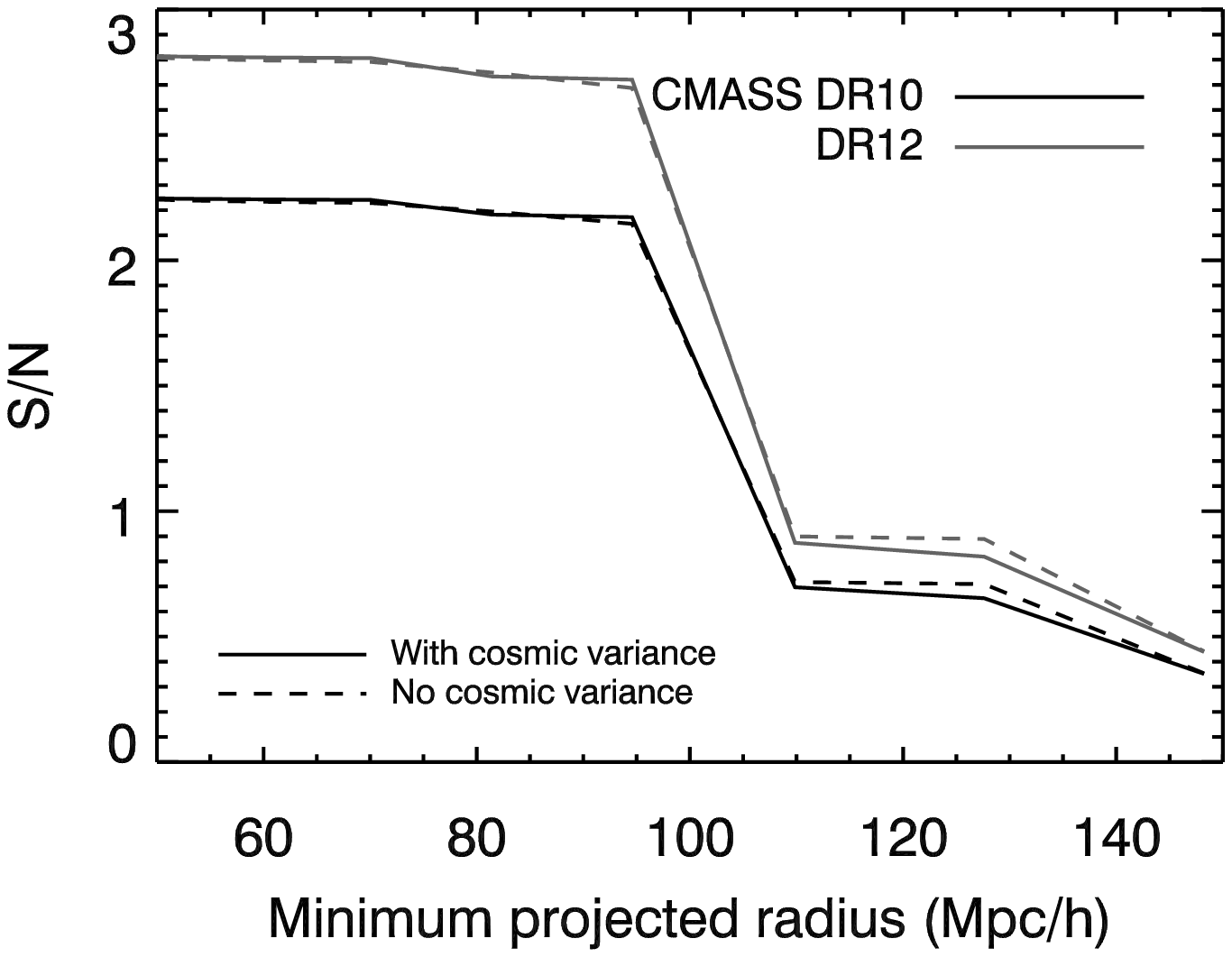}
\label{fig:SNbaocmass}
}
\subfigure[\hskip 2pt DESI]{
\includegraphics[width=0.45\textwidth]{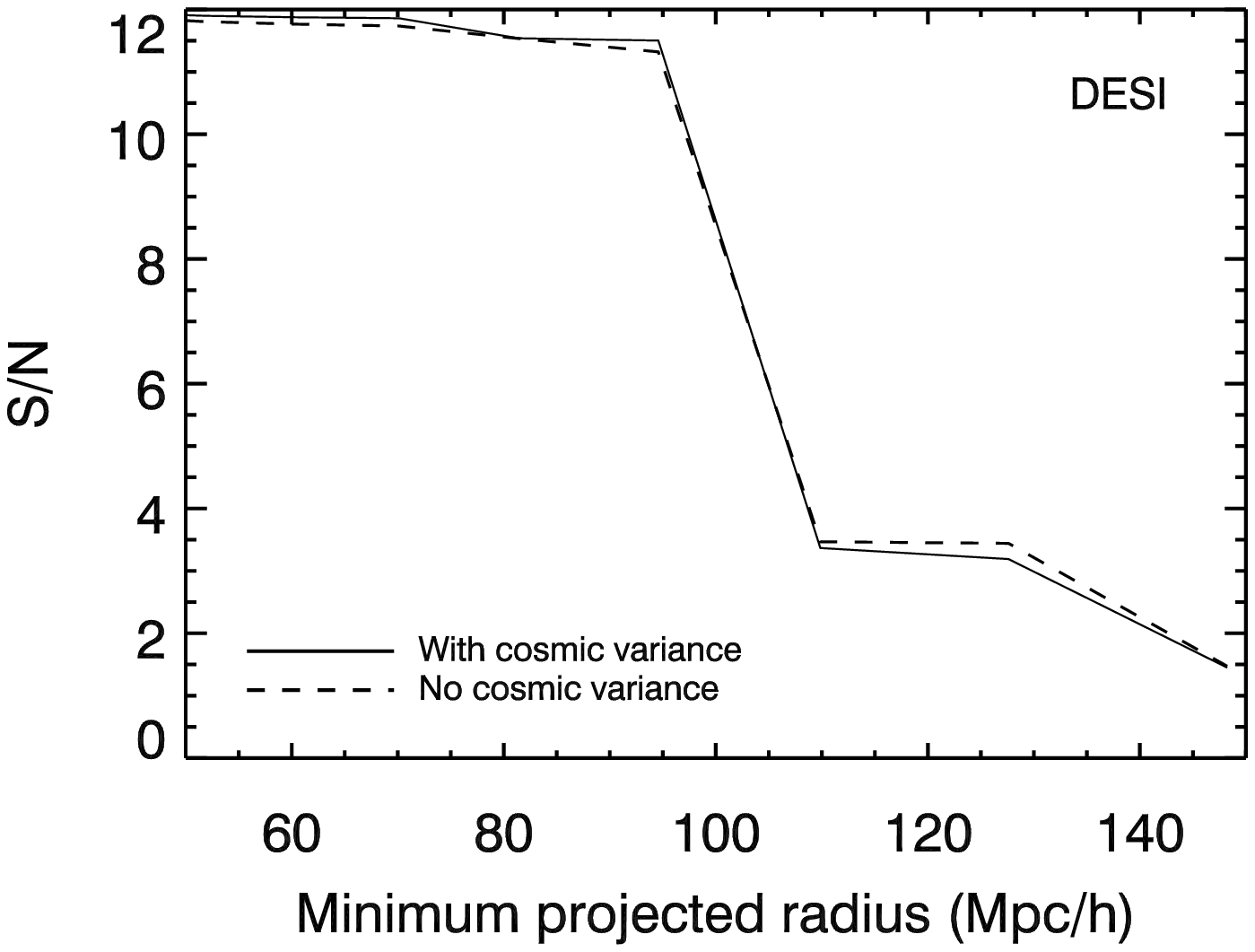}
\label{fig:snbaodesi}
}
\subfigure[\hskip 2pt EUCLID]{
\includegraphics[width=0.45\textwidth]{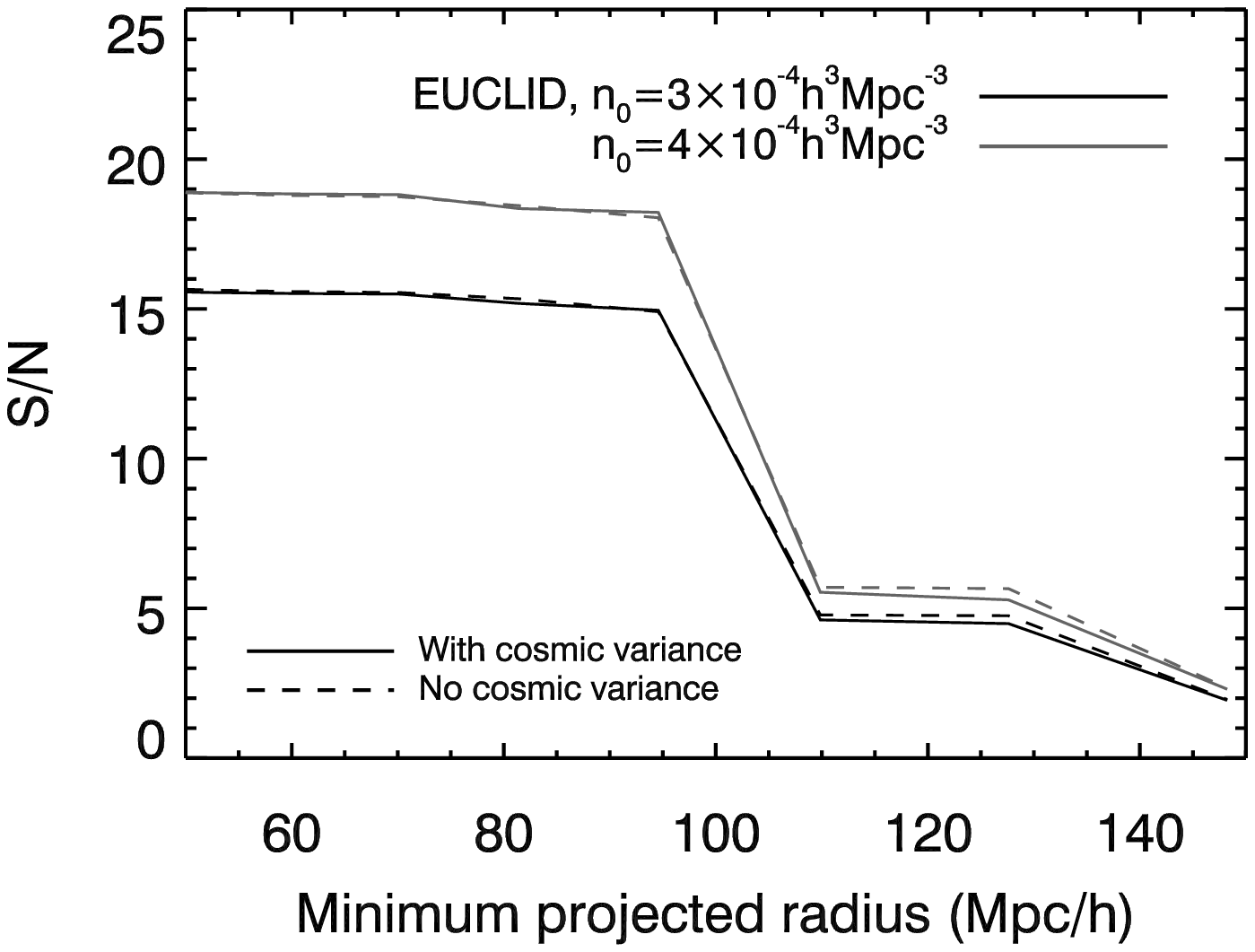}
\label{fig:snbaoEUCLID}
}
\caption{The signal-to-noise ratio of the BAO feature in the $w_{g+}$ cross-correlation as a function of the lower bound of the interval of projected radius for which it is calculated, keeping the upper bound fixed at $200$Mpc$/h$. We see a clear increase in the $S/N$ when the interval comprises the scale of the BAO, $\sim 110$Mpc$/h$. The $S/N$ converges to $\SNbaolowzTWELVE,\SNbaocmassTWELVE,\SNbaodesi$ for LOWZ DR12(\ref{fig:snbaolowz}), CMASS DR12 (\ref{fig:SNbaocmass}) and DESI (\ref{fig:snbaodesi}), respectively, at $r_p=80$ Mpc$/h$. For EUCLID (\ref{fig:snbaoEUCLID}), the $S/N$ predicted depends on the assumed comoving number density normalization, yielding $S/N=\SNbaoEUCLID$ and $S/N=\SNbaoEUCLIDfour$ for $n_0=3\times10^{-3}h^3$Mpc$^{-3}$ and $n_0=4\times10^{-3}h^3$Mpc$^{-3}$, respectively.}
\label{fig:baoSN}
\enf

\subsection{Comparison to galaxy-galaxy lensing}
\label{sec:galaxy-galaxy}

The detectability of the BAO and primordial non-Gaussianity in the galaxy-galaxy lensing signal ($gG$) were studied by \cite{Jeong09}. In the case of primordial non-Gaussianity, \cite{Jeong09} obtain the following expression for the projected surface mass density profile of a lens at $z_L$:

\beeq
\Delta\Sigma(r_p,z_L) = \rho_0 \int \frac{kdk}{2\pi} [b+\Delta b(k,z_L)]P_\delta(k,z_L)J_2(kr_p),
\label{eq:dsigma_jeong}
\eneq

\noindent where $\rho_0=2.77\times 10^{11}(\Omega_Mh^2)$M$_\odot$Mpc$^{-3}$ is the mean comoving mass density. This expression takes into account the scale dependent bias of the lenses in a similar spirit as our Eq. \eqref{eq:ngwgp}. The main difference between Eq. \eqref{eq:ngwgp} and Eq. \eqref{eq:dsigma_jeong} is the effect of the scale-dependent bias on RSD in the case of IA. 
We compare in Figure \ref{fig:gglens} the fractional difference in $\Delta\Sigma(r_p,z_L)$ and in $w_{g+}$ when non-Gaussianity of $\fnl=10$ is considered (similar results for $gG$ are presented in Figure 5 of \cite{Jeong09}). In the galaxy-galaxy lensing case, we choose $z_L$ and the bias values at the median redshift of the surveys considered. For LOWZ and CMASS, the bias is fixed at $b=2$, while for DESI and EUCLID, we scale the bias keeping the correlation length fixed and normalizing it to match the bias derived by \cite{Blazek11} for the sample of \cite{Okumura09} at $\bar{z}=0.32$. The fractional change in the inferred surface mass density profile and $w_{g+}$ are similar because they result from the scale-dependent bias. The effect of non-Gaussianity also enters $w_{g+}$ through the RSD factor. 

\cite{Jeong09} find that, for lens samples of LRGs and clusters at $z=\{0.3,0.5,0.8\}$, they cannot distinguish between $\fnl=0$ and $\fnl=100$ but they can achieve a significant detection of the BAO feature in the projected surface mass density from galaxy-galaxy lensing. As suggested in the previous section, $gG$ and $gI$ could be combined to yield a more robust detection of the BAO.

\bef
\centering
\includegraphics[width=0.5\textwidth]{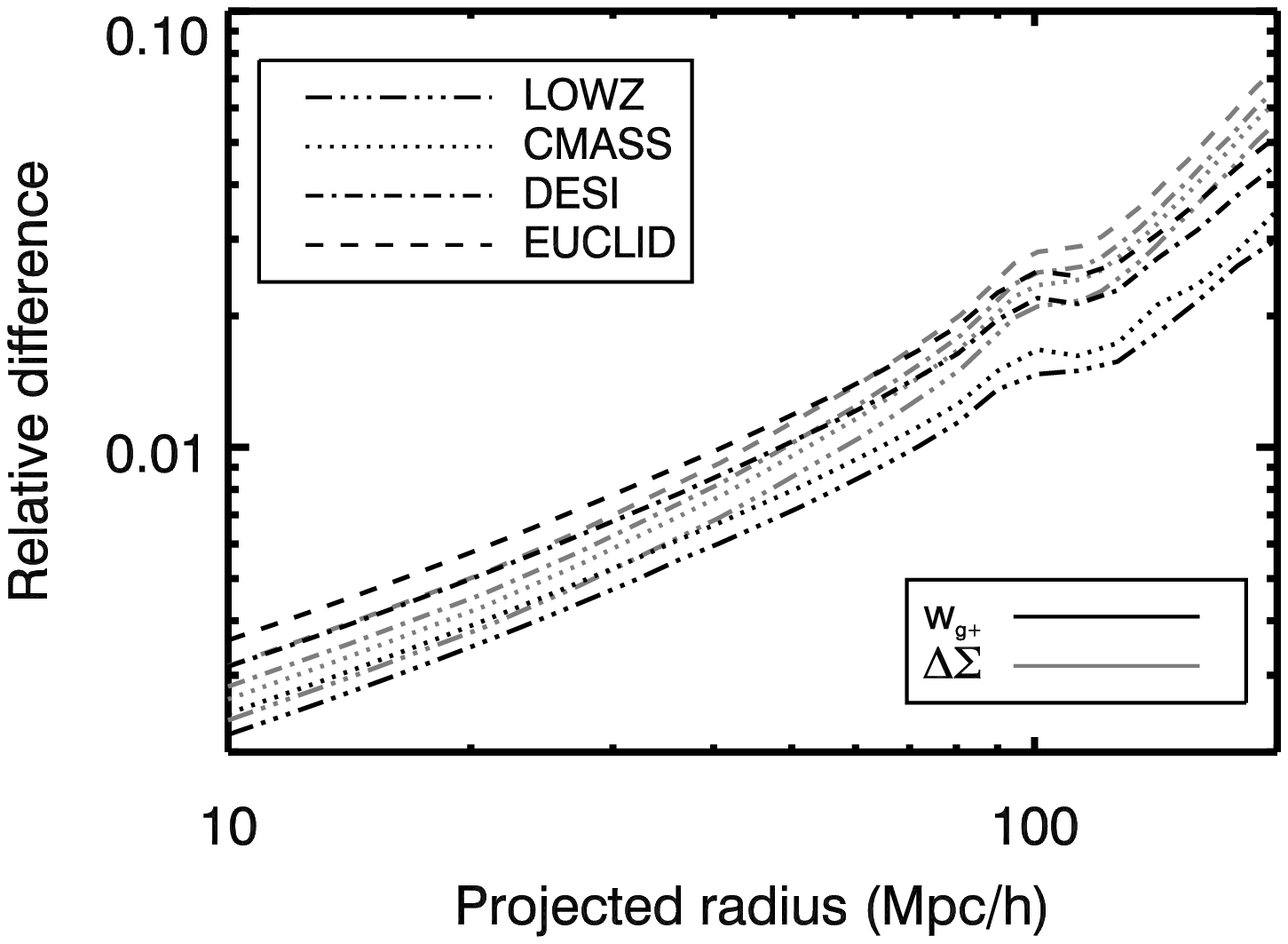}
\caption{The fractional effect of primordial non-Gaussianity ($\fnl=10$) and the BAO feature in the projected surface mass density profile obtained from galaxy-galaxy lensing for lenses at the median redshifts of LOWZ, CMASS, DESI and EUCLID LRGs.}
\label{fig:gglens}
\enf

\section{Conclusions}

The intrinsic alignments of LRGs may become a cosmological tool in the near future. We have presented forecasts for the detection of the BAO signature and of primordial non-Gaussianity in the cross-correlation function of LRG positions and shapes for ongoing and upcoming surveys. Our predictions are based on the LA model of \cite{Catelan01} and further development by \cite{Hirata04,Hirata10} and the NLA model of \cite{Bridle07}, which yield similar results on scales above $10$ Mpc$/h$ but differ on non-linear scales. Our results are expressed in terms of the projected correlation function of galaxy positions and intrinsic shears, $w_{g+}$. 

At low redshift, we constrain the strength of alignment, parametrized by $C_1$, by means of the observational results of \cite{Okumura09} for $w_{g+}$. Our result is consistent with that of \cite{Blazek11}. We explore the effect of smoothing the tidal field on different scales and we find that it has a significant impact on non-linear scales in the cross-correlation function of positions and shapes. The observational constraints are consistent with a smoothing over the scales of a typical LRG halo. The impact of the smoothing filter is smaller on small scales and larger on large scales in the auto-correlation functions of intrinsic shears compared to $w_{g+}$. While the NLA model reproduces the observed $w_{g+}$ surprisingly well even below $10$ Mpc$/h$, there is a need for simulations to study the alignments of LRGs on small scales to fully address non-linear physics. We also explored the redshift dependence of the LA model and compared the amplitude of $w_{g+}$ if alignments are fixed at the redshift of formation of a galaxy to the case where the galaxy reacts instantaneously to the large-scale tidal field. We have shown that at the current level of uncertainty, it is not possible to distinguish between the two models.

We computed the covariance matrix associated to $w_{g+}$, presented in the Appendix, and we have shown that while this matrix is dominated by shape noise on small scales, cosmic variance has a significant contribution on scales $>100$ Mpc$/h$ for all surveys considered. Cosmic variance is negligible in the scale of the BAO for the assumed value of $\sigma_\gamma$ except for the LOWZ sample, which probes a much smaller cosmological volume.

The BAO feature in $w_{g+}$ is different from the usual bump in the galaxy correlation function, $w_{gg}$. In the case of $w_{g+}$, the BAO appears as a consecutive trough, node and bump around the scale of $110$ Mpc$/h$, similarly to the observed effect on the $gG$ correlation function \cite{Jeong09}. The trough coincides with the position of the peak in the galaxy correlation function, at $\sim 100$ Mpc$/h$ for our fiducial cosmology. For the LOWZ and CMASS samples, the BAO detection would be marginally significant at $S/N = \SNbaolowzTWELVE$ and $S/N = \SNbaocmassTWELVE$ once BOSS is completed. For DESI and EUCLID, we obtain forecasts of significant detections at $S/N = \SNbaodesi$ (DESI), $\SNbaoEUCLID$ (EUCLID with $n_0=3\times10^{-3}h^3$Mpc$^{-3}$) and $\SNbaoEUCLIDfour$ (EUCLID with $n_0=4\times10^{-3}h^3$Mpc$^{-3}$).

We find that there are two non-Gaussian contributions to $w_{g+}$ to order $\mathcal{O}(\fnl)$. One of the terms comes from the non-Gaussian bias of large-scale structure \cite{Dalal08}. The second term is constant with scale but dependent on redshift. Interestingly, this term propagates the effect of non-Gaussianity from large to small scales. Unfortunately, it is at least $2$ orders of magnitudes below the contribution of the non-Gaussian bias term. We have also considered the effect of non-Gaussianity on RSD, as in \cite{Schmidt10}. Overall, we find that a value of $\fnl = 10$ (at the 1.2 $\sigma$ level with $Planck$ measurements) yields $S/N < 2$ for all surveys. 

On very large scales, above the scale of the BAO, the effect of RSD is roughly proportional to the effect of non-Gaussianity in projection. While in $k-$ space these terms have a different dependence, in projection, RSD can mimic primordial non-Gaussianity of the local type. Neglecting RSD in the modeling of $w_{g+}$ can lead to a spurious detection of primordial non-Gaussianity.

\acknowledgments
We thank Rachel Mandelbaum, Jonathan Blazek, Fabian Schmidt, Michael Strauss, and Matias Zaldarriaga for useful discussions. We also thank Teppei Okumura for providing us with his results on the cross-correlation function of LRG positions and their shapes based on data from the Sloan Digital Sky Survey. C.D. is supported by the National Science Foundation grant number AST-0807444, NSF grant number PHY-0855425, and the Raymond and Beverly Sackler Funds. 

Funding for SDSS-III has been provided by the Alfred P. Sloan Foundation, the Participating Institutions, the National Science Foundation, and the U.S. Department of Energy Office of Science. The SDSS-III web site is http://www.sdss3.org/.

SDSS-III is managed by the Astrophysical Research Consortium for the Participating Institutions of the SDSS-III Collaboration including the University of Arizona, the Brazilian Participation Group, Brookhaven National Laboratory, Carnegie Mellon University, University of Florida, the French Participation Group, the German Participation Group, Harvard University, the Instituto de Astrofisica de Canarias, the Michigan State/Notre Dame/JINA Participation Group, Johns Hopkins University, Lawrence Berkeley National Laboratory, Max Planck Institute for Astrophysics, Max Planck Institute for Extraterrestrial Physics, New Mexico State University, New York University, Ohio State University, Pennsylvania State University, University of Portsmouth, Princeton University, the Spanish Participation Group, University of Tokyo, University of Utah, Vanderbilt University, University of Virginia, University of Washington, and Yale University.

\bibliography{chisari_dvorkin}

\appendix
\section{Covariance Matrix Derivation}
\label{sec:appendix}

The correlation function $\xi_{g+}({\bf r_p},\Pi,z)$ presented in Eq. (\ref{eq:basicxi}) is not spherically symmetric and $w_{g+}$, in Eq. (\ref{eq:wgp}), represents a projection of $\xi_{g+}({\bf r_p},\Pi,z)$ along the direction of the line of sight, $\Pi$, between $-\Pi_{\rm max}$ and $\Pi_{\rm max}$, and an angular average on the plane of the sky. We present in this Appendix the calculation of the covariance matrix between radially averaged bins of the projected correlation function of two observables A and B in the case in which the spherical symmetry has been broken. We note that this is relevant, for example, in the presence of RSD. In the case of spherically symmetric correlation functions, a derivation of the covariance matrix is given by \cite{Smith09}.

The projected correlation function of two observables A and B, averaged over a radial bin with boundaries $[r_{i,\rm min},r_{i,\rm max}]$ and centered on $r_i$, is given by

\beeq
\bar{w}_{\rm AB}(r_i) = \frac{2\pi}{A_i}\int_{r_{i,\rm min}}^{r_{i,\rm max}} dr\, r\, w_{\rm AB}(r),
\eneq

\noindent where $w_{\rm AB}(r)$ is the continuous projected correlation function, and $A_i$ is the area of bin $i$.

For computing the covariance matrix element between two radial bins $i$ and $j$ of the projected correlation function, we first compute the covariance matrix of the power spectrum of observables A and B,

\beeq
C_{P_{AB}}^d = \langle P^d_{AB}({\kbf_1},z) P^d_{AB}({\kbf_2},z)\rangle - \langle  P^d_{AB}({\kbf_1},z)\rangle  \langle  P^d_{AB}({\kbf_2},z)\rangle,
\eneq

\noindent where the superscript $d$ indicates that the observables are discrete, i.e., the galaxy density field is not a continuous field, but is rather sampled at the positions of the galaxies. For the density-intrinsic shear power spectrum,

\bear
C_{P_{g+}}^d(\kbf_1,\kbf_2) &=& \frac{(2\pi)^3\delta^{(D)}(\kbf_1+\kbf_2)}{V_s} \left[ P_{gg}(\kbf_1,z) + \frac{1}{n(z)}\right]\left[P_{++}(\kbf_1,z)+ \frac{\sigma_\gamma^2}{n(z)}\right]\nonumber\\
&+&\frac{(2\pi)^3\delta^{(D)}(\kbf_1-\kbf_2)}{V_s}  P_{g+}^2(\kbf_1,z),
\enar

\noindent where $\sigma_\gamma$ is the intrinsic scatter in the galaxy shears, $n(z)$ is the average comoving number density of galaxies at a given redshift and $V_s$ is the survey volume. 

We work in cylindrical coordinates both in Fourier space and in real space; the components of the $\kbf$ and ${\bf r}$ vectors are $(k_{\perp},\theta,k_z)$ and $(r_p,\phi,\Pi)$ respectively. The angle between ${\bf r}_1$ and ${\bf r}_2$ is $\phi_{12}$. The covariance element of the correlation $w_{\rm AB}$ for radii $r_1$ and $r_2$ can be obtained by a Fourier transform of $C_{P_{AB}}^d $,

\bear
C_{w_{AB}}(r_1,r_2)&=& (2\Pi_{\rm max})^2  \int dz \mathcal{W}^2(z)  \int \frac{d^3\kbf_1}{(2\pi)^3} \int \frac{d^3\kbf_2}{(2\pi)^3}C_{P_{AB}}^d({\bf k_1},{\bf k_2},z)\nonumber\\
&\times& j_0(k_{1z} \Pi_{\rm max})j_0(k_{2z} \Pi_{\rm max})e^{i{\bf k_{1\perp}\cdot r_1}}e^{i{\bf k_{2\perp}\cdot r_2}},
\label{covwab}
\enar 

\noindent where we have applied the Limber approximation consistently with our computation of the projected correlation function in Section \ref{sec:correlation}. 

To obtain the covariance matrix of the bin averaged correlation, we average over the annuli,

\bear
C_{\bar{w}_{AB}}(r_i,r_j)&=& \frac{(2\Pi_{\rm max})^2}{A_iA_j} \int_{A_i}  d^2r_1\int_{A_j}  d^2r_2 \int dz \mathcal{W}^2(z)  \int \frac{d^3\kbf_1}{(2\pi)^3} \int \frac{d^3\kbf_2}{(2\pi)^3}C_{P_{AB}}^d({\bf k_1},{\bf k_2},z)\nonumber\\
&&\times j_0(k_{1z} \Pi_{\rm max})j_0(k_{2z} \Pi_{\rm max})e^{i{\bf k_{1\perp}\cdot r_1}}e^{i{\bf k_{2\perp}\cdot r_2}}.
\label{covwabbar}
\enar 

In the density-intrinsic shear case, it follows that Eq. (\ref{covwab}) can be specified to be

\bear
C_{\bar{w}_{g+}}(r_i,r_j)&=& \frac{(2\Pi_{\rm max})^2}{(2\pi)^3V_sA_iA_j}  \int dz \mathcal{W}^2(z) \int_{A_i}  d^2r_1\int_{A_j}  d^2r_2  \int d^3\kbf_1 \int d^3\kbf_2\, e^{i{\bf k_{1\perp}\cdot r_1}}e^{i{\bf k_{2\perp}\cdot r_2}}\nonumber\\
&\times& j_0(k_{1z} \Pi_{\rm max})j_0(k_{2z} \Pi_{\rm max})\times\nonumber\\
&& \{\delta^{(D)}(\kbf_1+\kbf_2) \left[ P_{gg}(\kbf_1,z) + n^{-1}(z)\right]\left[P_{++}(\kbf_1,z)+\sigma_\gamma^2 n^{-1}(z)\right] \nonumber\\
&+&\delta^{(D)}(\kbf_1-\kbf_2) P_{g+}^2(\kbf_1,z)\},
\enar 

\noindent where $A_i$ and $A_j$ are the areas of the radial bins over which the average of the correlation function is performed, delimited by $[r_{i,\rm min},r_{i,\rm max}]$ and $[r_{j,\rm min},r_{j,\rm max}]$. Performing the integration for each term yields the following expression for the covariance matrix,

\bear
C_{\bar{w}_{g+}}(r_i,r_j)
&=&  \frac{4\pi\Pi_{\rm max}\mathcal{F}(r_i,r_j,\Delta r)}{V_s A_j A_i} \ \int dz \mathcal{W}^2(z) \frac{\sigma_\gamma^2}{n(z)^2}
+ \frac{(2\Pi_{\rm max})^2}{V_sA_iA_j}  \frac{b^2\sigma_\gamma^2}{n(z)} \int dz \mathcal{W}^2(z)\nonumber\\ 
&\times&\int dk_{1z}{dk_{1\perp}\over k_{1\perp}} j_0^2(k_{1z} \Pi_{\rm max}) P_{\delta}(\kbf_1,z)\times\nonumber\\
&&\left[r_{j,\rm max}j_1\left(k_{1\perp}r_{j,\rm max}\right)- 
r_{j,\rm min}j_1\left(k_{1\perp}r_{j,\rm min}\right)\right]\left[r_{i,\rm max}j_1\left(k_{1\perp} r_{i,\rm max}\right)- 
r_{i,\rm min} j_1\left(k_{1\perp} r_{i,\rm min}\right)\right]+\nonumber\\
&+&\frac{\Pi_{\rm max}^2}{\pi V_sA_iA_j}\int_0^{2\pi} d\phi_{12}\int_{i}  dr_1\,r_1\int_{j}  dr_2\,r_2   \int dz \mathcal{W}^2(z) \left(\iacons\right)^2\nonumber\\
&\times&\int dk_{1z}dk_{1\perp} \frac{k_{1\perp}^5}{k_1^4} j_0^2(k_{1z} \Pi_{\rm max}) \mathcal{S}^2(k_{1\perp})\mathcal{S}^2(k_{1z})P_{\delta}(\kbf_1,z)\nonumber\\
&\times&\{\left[b^2P_{\delta}(\kbf_1,z)+ n^{-1}(z) \right] \left[J_0(k_{1\perp}|{\bf r_1}-{\bf r_2}|)+J_4(k_{1\perp}|{\bf r_1}-{\bf r_2}|)\right]\nonumber\\
&+& b^2 P_{\delta}(\kbf_1,z) \left[J_0(k_{1\perp}|{\bf r_1}+{\bf r_2}|)+J_4(k_{1\perp}|{\bf r_1}+{\bf r_2}|)\right]\},
\label{fullcov}
\enar 

\noindent where $j_0(x)$ is the zeroth order spherical Bessel function, $J_0$ and $J_4$ are Bessel functions of the first kind, and we have defined the functions $\mathcal{J}_1$, 

\bear
\mathcal{J}_1(a,b) &\equiv& ab\,\int dx j_1(ax)j_1(bx)x^{-1} = \frac{ab(a^2+b^2)-(a^2-b^2)^2}{8ab}{\rm atanh}\left(\frac{b}{a}\right),
\enar

\noindent and $\mathcal{F}(r_i,r_j,\Delta r)$,

\bear
\mathcal{F}(r_{i,\rm min},r_{i,\rm max},r_{j,\rm min},r_{j,\rm max}) &\equiv& \mathcal{J}_1\left(r_{i,\rm max},r_{j,\rm max}\right)- \mathcal{J}_1\left(r_{i,\rm max},r_{j,\rm min}\right)\nonumber\\
&+&\mathcal{J}_1\left(r_{i,\rm min},r_{j,\rm min}\right)-\mathcal{J}_1\left(r_{j,\rm max},r_{i,\rm min}\right).
\enar

\noindent For each power of $P_\delta$ in Eq. (\ref{fullcov}), there is a factor due to redshift space distortions in the Kaiser approximation, $\left(1+\beta k_z^2/k^2\right)^2$, that we have left implicit. As mentioned in Section \ref{sec:observed}, $\beta$ is related to the growth factor through $\beta=\Omega_M^{0.55}(z)/b(z)$.

Figure \ref{fig:cc_CMASS} shows the correlation coefficient for the covariance matrix of $\bar{w}_{g+}(r_i,r_j)$ for the CMASS sample in DR10 assuming the comoving number density shown in Figure \ref{fig:comovn}. On small scales, the covariance matrix is predominantly diagonal due to the dominance of the shape noise term. Figure \ref{fig:ccratio_CMASS} shows the ratio between the covariance matrix that only considers shape noise (the first two terms of Eq. \ref{fullcov}), compared to the full covariance matrix, including cosmic variance terms. The correlation coefficient for DESI and EUCLID are also shown in Figures \ref{fig:cc_DESI} and \ref{fig:cc_EUCLID}, respectively. We also compare the full covariance matrix to the shape noise-only covariance in Figures \ref{fig:ccratio_DESI} and \ref{fig:ccratio_EUCLID}. For CMASS, DESI and EUCLID, cosmic variance terms only become significant at scales of above $100$Mpc$/h$. The LOWZ DR10 sample, in the contrary, shows a significant effect from cosmic variance on small scales in Figures \ref{fig:cc_LOWZ} and \ref{fig:ccratio_LOWZ} due to the small cosmological volume probed. In general, cosmic variance does not affect the detection of the BAO except for the LOWZ sample (both in DR10 and DR12), but it can have a significant impact on the detection of primordial local non-Gaussianity.

Convergence tests of the computation of the covariance matrix were performed for all terms that required numerical integration in Eq. (\ref{fullcov}) and in all variables ($k_z$, $k_{\perp}$, $z$, $r_1$, $r_2$, and $\phi_{12}$). The level of convergence achieved in all of these cases was always below $7\%$.

\bef
\centering
\subfigure[\hskip 2pt Correlation coefficient $r_{ij} = \frac{C_{ij}}{\sqrt{C_{ii}C_{jj}}}$ for the covariance matrix of $\bar{w}_{g+}(r_i,r_j)$ with logarithmic spacing for the radial bins.]{
\includegraphics[width=0.45\textwidth]{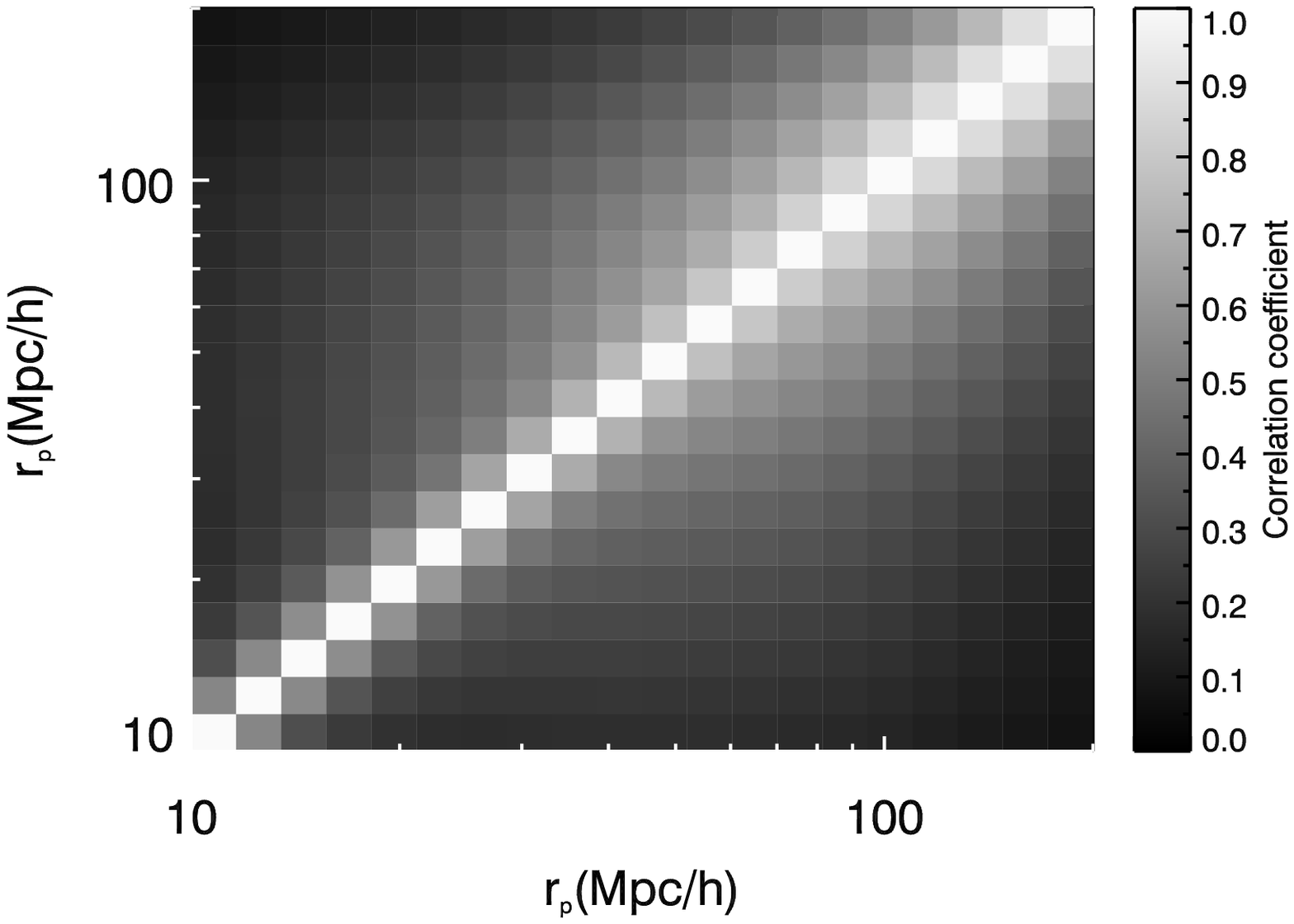}
\label{fig:cc_LOWZ}
}
\subfigure[\hskip 2pt The ratio between the shape noise-only covariance matrix and the full covariance matrix.]{
\includegraphics[width=0.45\textwidth]{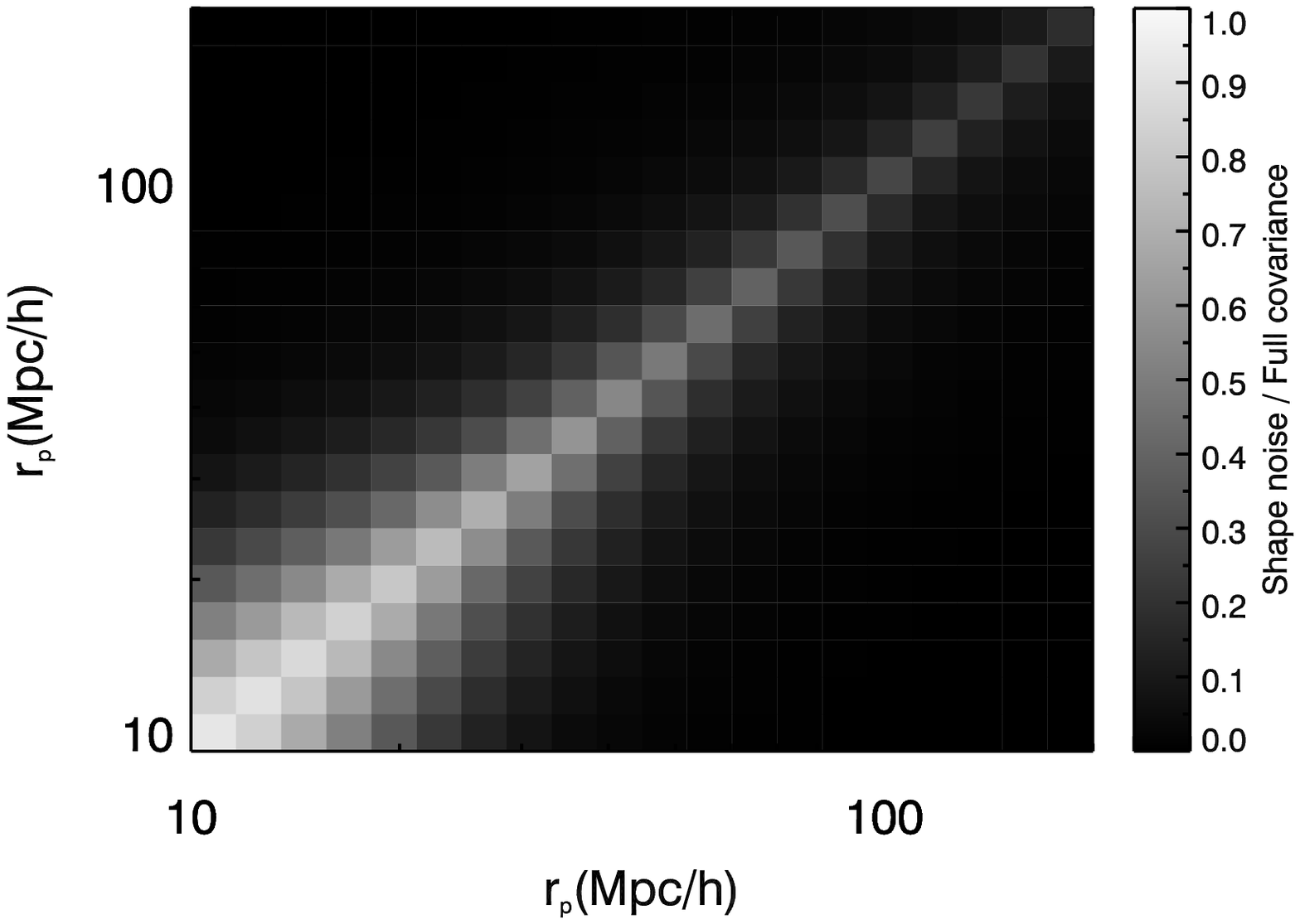}
\label{fig:ccratio_LOWZ}
}
\caption{LOWZ DR10.}
\enf

\bef
\centering
\subfigure[\hskip 2pt Correlation coefficient $r_{ij} = \frac{C_{ij}}{\sqrt{C_{ii}C_{jj}}}$ for the covariance matrix of $\bar{w}_{g+}(r_i,r_j)$ with logarithmic spacing for the radial bins.]{
\includegraphics[width=0.45\textwidth]{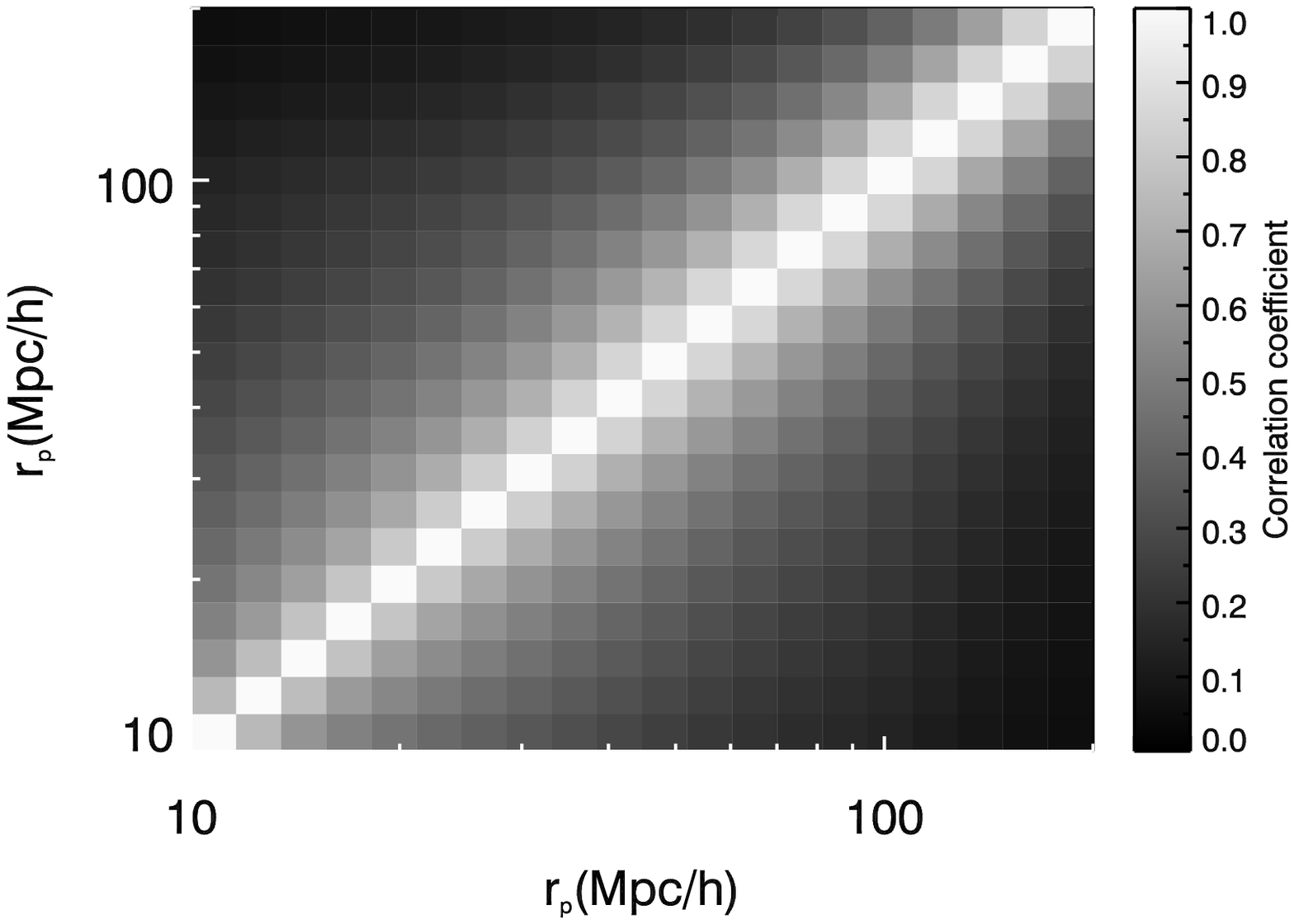}
\label{fig:cc_CMASS}
}
\subfigure[\hskip 2pt The ratio between the shape noise-only covariance matrix and the full covariance matrix.]{
\includegraphics[width=0.45\textwidth]{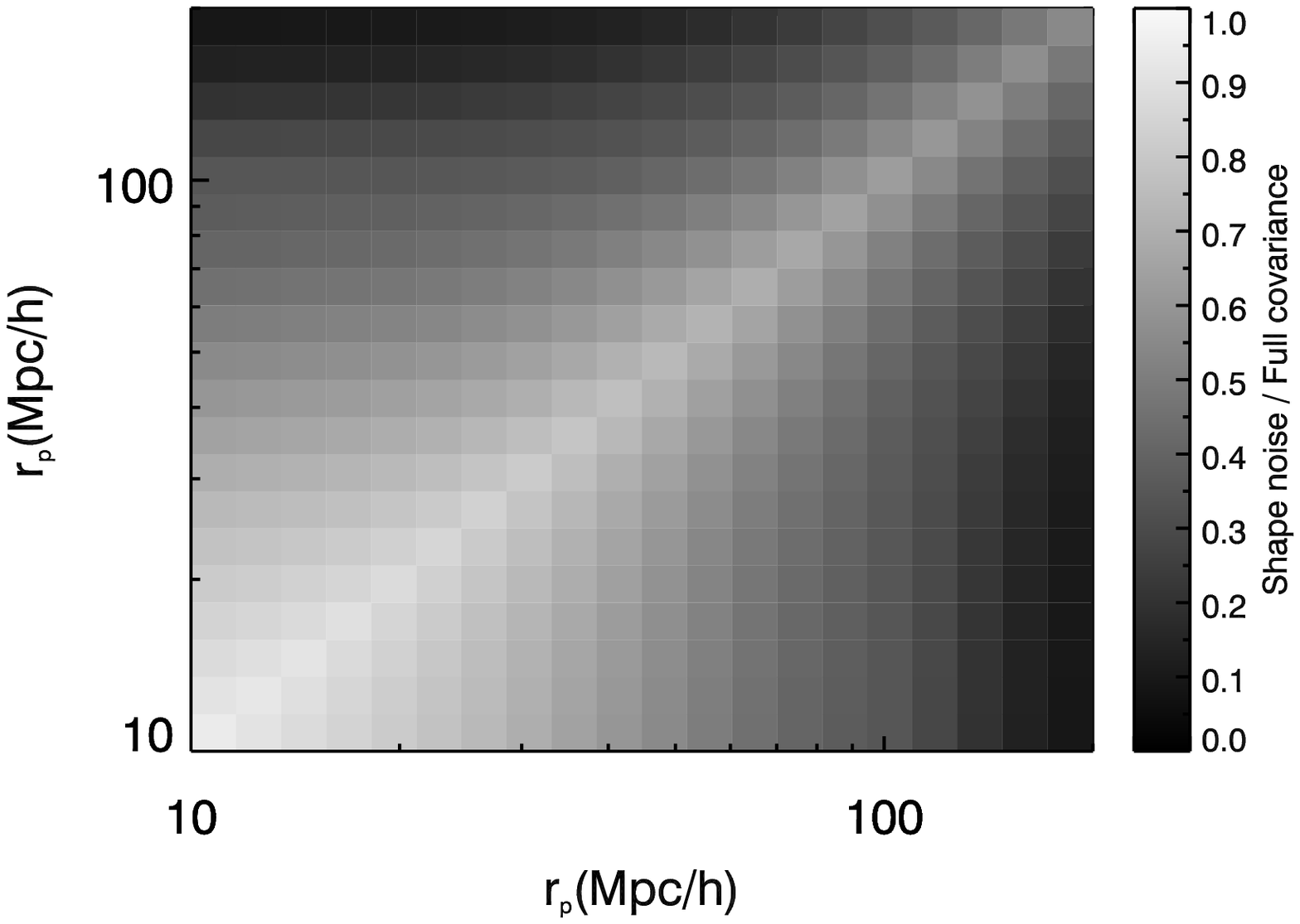}
\label{fig:ccratio_CMASS}
}
\caption{CMASS DR10.}
\enf

\bef
\centering
\subfigure[\hskip 2pt Correlation coefficient $r_{ij} = \frac{C_{ij}}{\sqrt{C_{ii}C_{jj}}}$ for the covariance matrix of $\bar{w}_{g+}(r_i,r_j)$ with logarithmic spacing for the radial bins.]{
\includegraphics[width=0.45\textwidth]{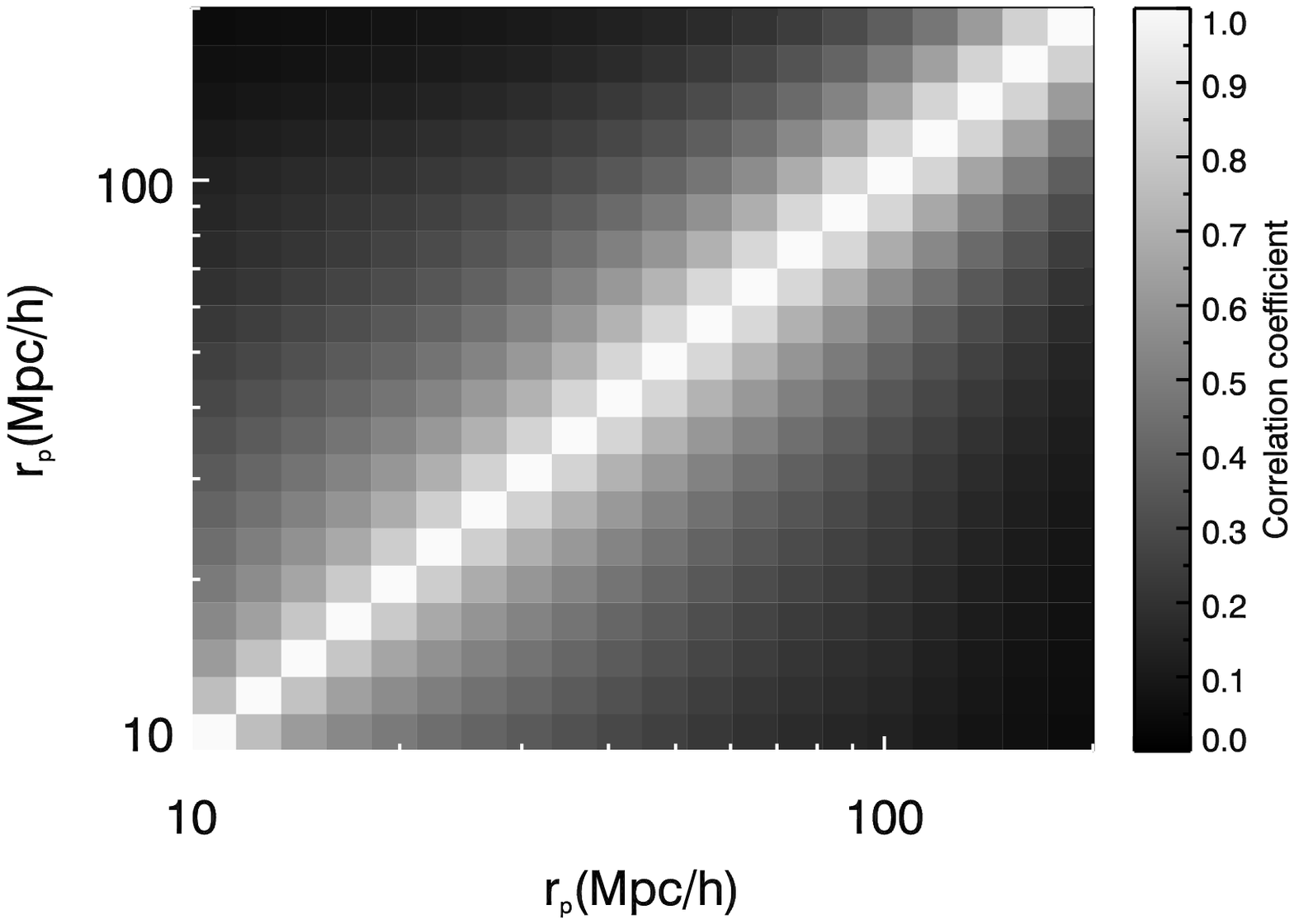}
\label{fig:cc_DESI}
}
\subfigure[\hskip 2pt The ratio between the shape noise-only covariance matrix and the full covariance matrix.]{
\includegraphics[width=0.45\textwidth]{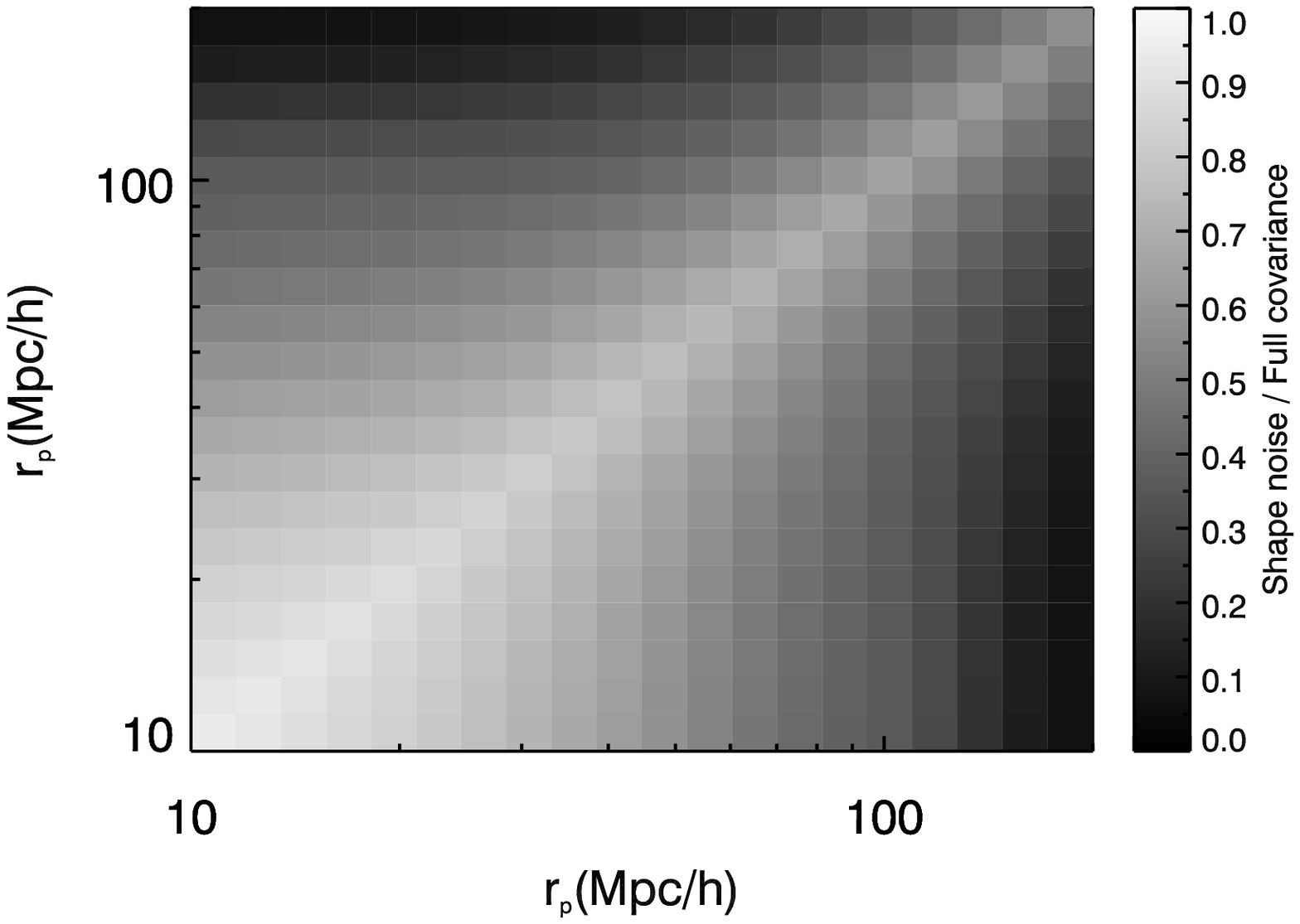}
\label{fig:ccratio_DESI}
}
\caption{DESI.}
\enf

\bef
\centering
\subfigure[\hskip 2pt Correlation coefficient $r_{ij} = \frac{C_{ij}}{\sqrt{C_{ii}C_{jj}}}$ for the covariance matrix of $\bar{w}_{g+}(r_i,r_j)$ with logarithmic spacing for the radial bins.]{
\includegraphics[width=0.45\textwidth]{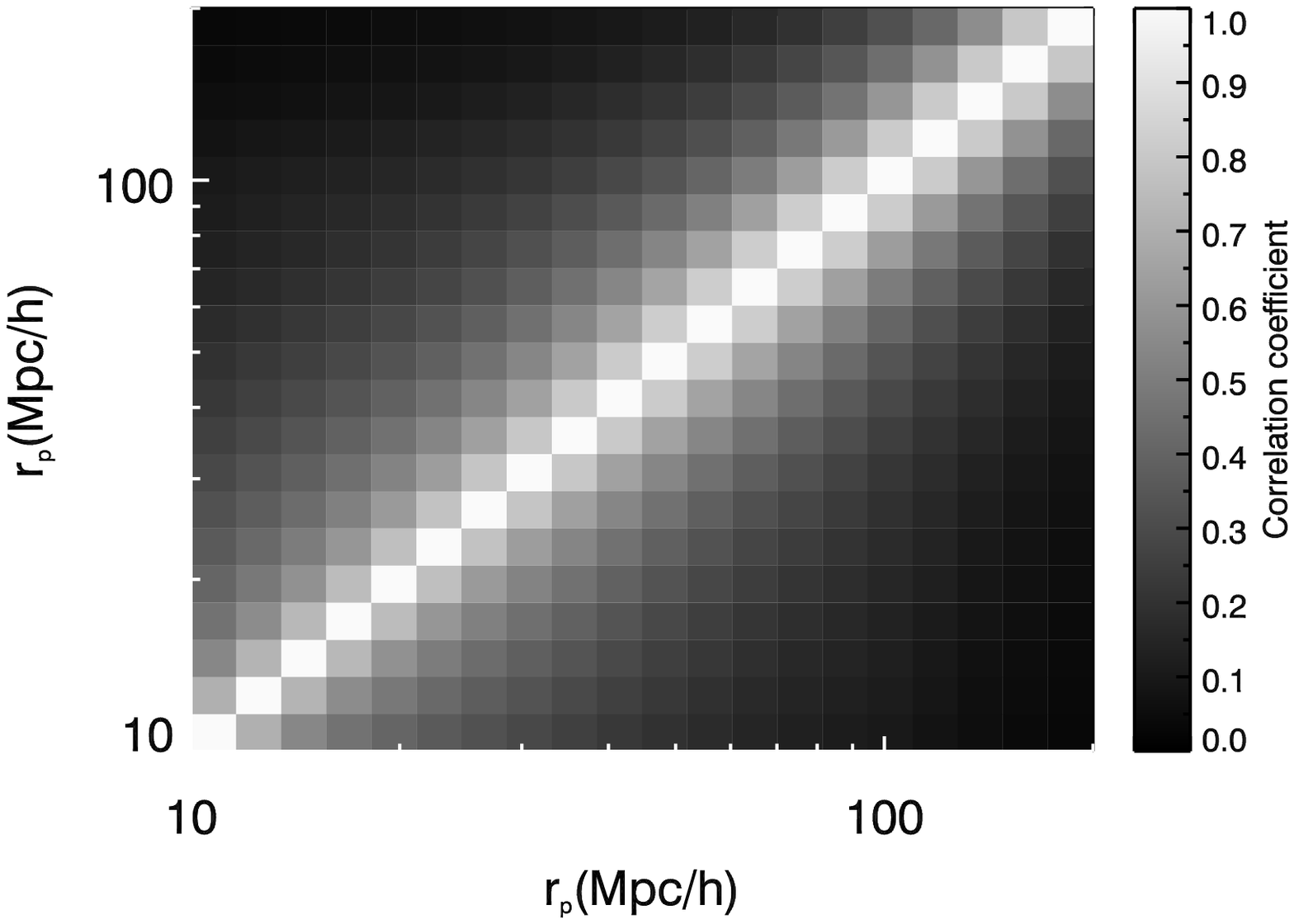}
\label{fig:cc_EUCLID}
}
\subfigure[\hskip 2pt The ratio between the shape noise-only covariance matrix and the full covariance matrix.]{
\includegraphics[width=0.45\textwidth]{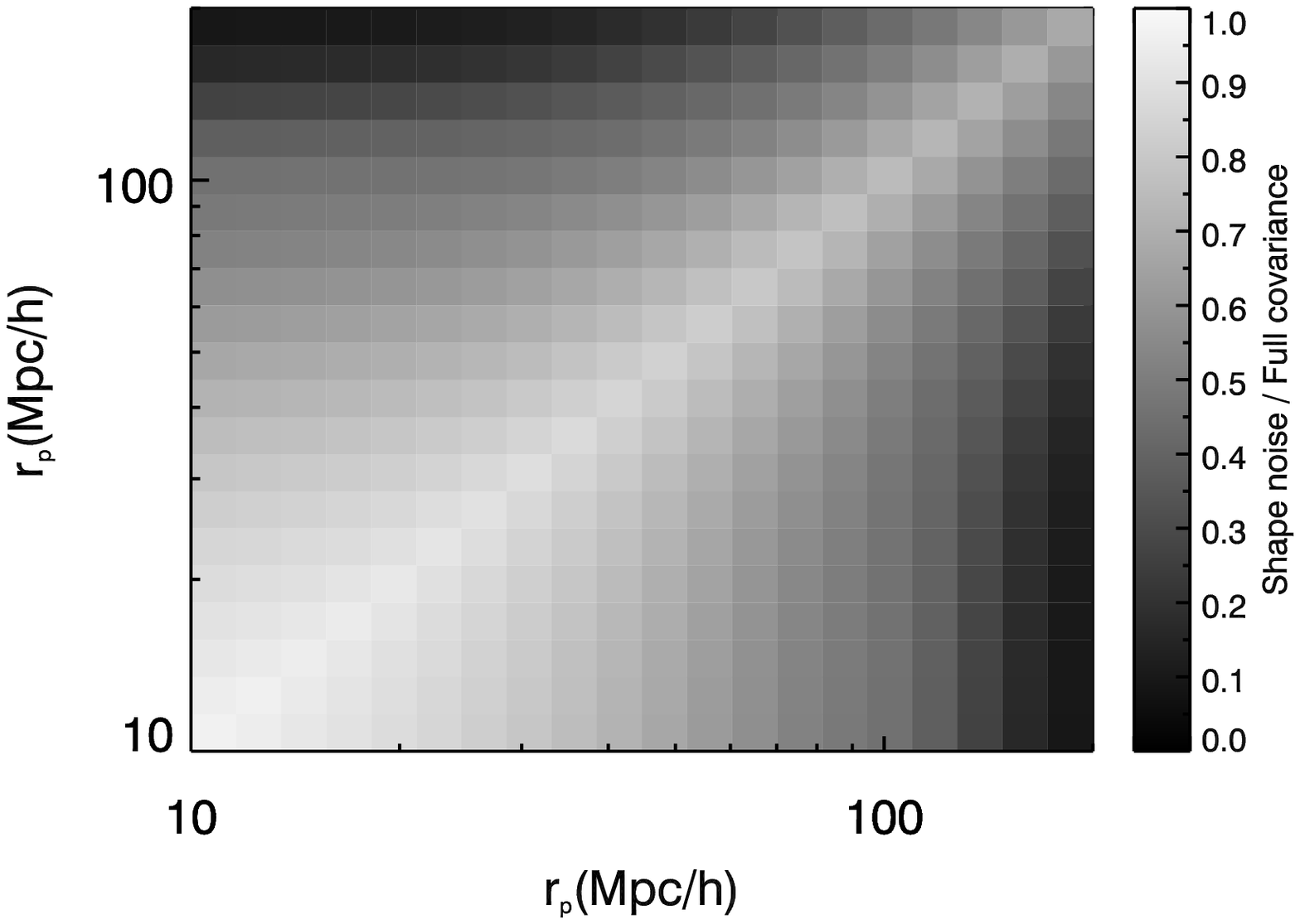}
\label{fig:ccratio_EUCLID}
}
\caption{EUCLID with $n_0 = 3\times10^{-4}h^3$Mpc$^{-3}$.}
\enf

\end{document}